# Embedded Polygon Symbolic Transfer Entropy (EPSTE): A Geometric Token and Deep Learning Approach to Estimating Transfer Entropy in Neuroimaging Time Series

David Finnigan

## Abstract

Inferring directed interactions between neural systems from EEG and MEG remains challenging due to noise, nonstationarity, and the high sample complexity of information-theoretic estimators. Transfer Entropy (TE) provides a principled and model-free measure of directed information flow, however its practical estimation is not stable in finite data regimes (particularly as embedding dimension increases). This work introduces Embedded Polygon Symbolic Transfer Entropy (EPSTE), a framework that reframes TE estimation as a learnable problem operating on structured symbolic representations of local temporal morphology rather than raw signal amplitudes. Neural time series are decomposed into sequences of geometric primitives derived from local triplets of samples encoding complementary aspects of waveform structure such as magnitude, curvature and directional change. These primitives are discretised into symbolic tokens, yielding a compact but expressive state space over which symbolic TE is estimated. A recurrent neural network with attention-based multiple-instance learning is trained to predict surrogate-validated TE values from bags of symbolic temporal windows. The method is evaluated on source-reconstructed MEG data parcellated using the AAL90 atlas and compared against a standard symbolic baseline using identical architectures and supervision. The results demonstrate that while local window-level predictions are noisy, aggregation across trials and channel pairs yields stable directed dependencies. At the pair level, EPSTE achieves near-perfect recovery of ground-truth directed structure (Pearson $r \approx 0.99$, $R^2 \approx 0.98$) and significantly lower absolute error than the baseline (Wilcoxon signed-rank test, $p \approx 2.9 \times 10^{-15}$), indicating that representational geometry plays a critical role in enabling practical learnability of information-theoretic dependencies.

## 1. Introduction

As neuroscience advances, so too must its methods for modelling the complex, dynamic relationships that underlie brain function. A central aim of contemporary neuroscience is to understand how neural systems interact causally to give rise to cognition, perception, and behaviour (Bergmann & Hartwigsen, 2021; Seth, 2007). Although modern neuroimaging techniques such as EEG (Electroencephalography) and MEG (magnetoencephalography) provide increasingly detailed measurements of neural activity, extracting meaningful causal structure remains a significant methodological challenge (Seth et al., 2015). Presently, correlational analysis are inherently limited because they characterise undirected co-fluctuations between signals and cannot distinguish driver–receiver relationships or recover the directional structure of information flow (Haufe et al., 2013). As a result, correlational approaches offer limited insight into the causal organisation of neural systems.

These limitations have motivated the development of causal inference frameworks, most notably Granger causality (GC) and Transfer Entropy (TE). Transfer Entropy was introduced by Schreiber 2000 in a paper titled: "Measuring information transfer". TE is an information theoretic measure intended to quantify directed and asymmetric information transfer between stochastic processes. Rather than relying on linear models TE evaluates whether knowledge of the past of one process improves prediction of another beyond what can be achieved from the target's own history alone (Vicente et al., 2010). Within this framework, TE has proven somewhat valuable for identifying directed functional interactions in EEG and MEG data (Seth et al., 2015).

However, despite its conceptual appeal, practical estimation of Transfer Entropy from empirical neural data remains challenging (Mölter & Goodhill, 2020). TE depends on conditional probability distributions that must be inferred from finite samples. This task is made difficult by the nonlinear, noisy, and nonstationary nature of neural time series as commonly represented by raw altitudinal readings. These challenges have motivated the development of symbolic formulations of TE, including Symbolic Transfer Entropy (STE) (Li et al., 2020) and others, such as Symbolic Phase Transfer Entropy (SPTE), in which symbols are defined using a low-dimensional embedding that captures the local direction of change between groups of successive samples (Zhang et al., 2017). While this representation improves tractability and enhances robustness to limited volumes of noise and non-stationarity, it yields a relatively limited symbolic alphabet, which can constrain expressive capacity of the symbolic state space, constraining the richness of causal structure that can be recovered (Tabor, 2009).

More recently, advances in deep learning have now opened new avenues for information transfer and causal inference in time-series data (Sun et al., 2024). Recurrent neural networks and related architectures provide flexible, nonlinear sequence models capable of learning complex temporal dependencies (Ohno & Kumagai, 2021). Neural approximations of information-theoretic quantities, including Transfer Entropy, have therefore emerged as a promising alternative to classic estimator-based methods. However, most existing learning-based approaches operate directly on raw amplitude time series or implicit probability distributions, often requiring large datasets and careful optimisation to remain stable (Sun et al., 2024). These considerations motivate the exploration of alternative representations that reduce estimation complexity while preserving the spatiotemporal and structural information central to Transfer Entropy. The central research question of this work is therefore:

*Can Transfer Entropy between neural time series be reliably approximated by a recurrent neural network operating on more geometrically complex symbolic representations, and how does the choice of embedding dimension affect this approximation?*

To address this question, this work investigates a novel data representation framework for improving learnability in deep learning networks geared towards approximating information transfer in a Symbolic Transfer Entropy (STE) paradigm. This framework is termed Embedded Polygon Symbolic Transfer Entropy (EPSTE). Rather than estimating TE through direct probability density estimation, EPSTE decomposes time series into geometric primitives derived from local triplets of points along the x-axis. These primitives encode the local morphological structure of the signal and are discretised into symbolic sequences. Geometric features from these symbolic representations are then used as inputs to a Recurrent Neural Network (RNN) with an attentional multi-instance-learning (MIL) framework trained on TE estimates. This framework enables systematic investigation of how symbolic encoding strategies and embedding dimensionality influence the learnability and stability of neural approximations to STE.

The following hypotheses are evaluated:

- $H0_1$: *Polygon-based symbolic encoding does not increase learnability more effectively than classical amplitude-based time series representations.*
- H1: *Polygon-based symbolic encoding increases learnability more effectively than classical amplitude-based time series representations.*

# 2. Literature Review

## 2.1 Informational Causality in Neural Systems

In the context of neuroimaging and information theory, causality is typically defined statistically (Solo, 2008; Bressler & Seth, 2011)) rather than by the notion of causality employed by ontological philosophy (Simchi, 2023). Instead, causality reflects temporal precedence and predictive utility, a statistical conception of causal influence grounded in information transfer. Informational causality here is a process where $X$ is said to exert a causal influence on a process $Y$ if knowledge of the past of $X$ improves prediction of the future of $Y$, beyond what can be achieved using the past of $Y$ alone (Bressler & Seth, 2011). This formulation is of particular relevance for neuroimaging in which direct intervention or manipulation of neural systems is often infeasible, and causal inference must instead rely on observed temporal dependencies between target regions (Kiebel et al., 2006).

This notion underpins modern approaches to causality in directed functional connectivity in EEG and MEG research. Informational causality provides a pragmatic framework for assessing directed influence between macroscopic neural signals (Bressler & Seth, 2011). As can be derived from that which is referenced above there are two key statistical paradigms for representing information transfer ( a proxy for causality) based approaches for deriving causal phenomena in statistical empiricism, Granger Causality and Transfer Entropy (the key paradigm in this study).

## 2.2 Issues in EEG and MEG Data Characteristics

There are several key data characteristics in EEG and MEG time series that provide challenges in analysis of information transfer. EEG and MEG provide non-invasive measurements of the electric and magnetic fields [respectively] with millisecond-scale temporal resolution, enabling investigation of fast electrophysiological dynamics and time-lagged interactions (Ahlfors & Mody, 2016). This temporal precision makes these modalities particularly attractive for studying directed neural interactions. Neural signals exhibit time-varying statistical properties driven by changes in cognitive state, task demands, and intrinsic neural dynamics (Sudre et al., 2012). These properties can easily lead to violations of some of the central assumptions for many classical causal inference methods and complicate reliable estimation of directed interactions. A further challenge arises from source mixing. Each sensor records a linear superposition of activity from multiple neural generators due to volume conduction and field spread. This instantaneous mixing introduces spurious correlations and obscures true directional relationships, particularly at the sensor level (Palva et al., 2017).

## 2.3 Granger Causality in EEG/MEG

Granger causality formalises a predictive notion of causality within a linear autoregressive framework and has been widely applied in neuroscience. In the special case of jointly Gaussian processes, Granger causality has been shown to be theoretically equivalent to Transfer Entropy. The applicability of Granger causality to EEG and MEG data is constrained by both modelling assumptions and measurement properties. Classical Granger causality relies on linear vector autoregressive models and assumes that the underlying processes are stationary over the analysis window. However, this equivalence holds under restrictive assumptions, including linearity, stationarity, and Gaussian noise conditions that are frequently violated in macroscopic readings of cortical data. Consequently, while Granger causality provides a useful conceptual foundation, its practical applicability to EEG and MEG is limited due to restrictions imposed on data (Barnett et al., 2009).

In contrast, EEG and MEG signals are highly nonlinear, generally exhibit strong nonstationary in higher order and larger scale neural system (Kaplan et al., 2005), and are contaminated by measurement noise and physiological artefacts (Bahners et al., 2023). These challenges are exacerbated in multichannel recordings, where volume conduction in EEG and field spread in MEG lead to instantaneous mixing of neural sources at the sensor level. This mixing introduces strong zero-lag correlations between channels, which can obscure or mimic time-lagged dependencies required for reliable directional inference (O'Neill et al., 2015). Due to constraints GC may yield unstable, bidirectional, or spurious causal estimates when applied directly to sensor-level data.

While various extensions of Granger causality have been proposed to mitigate these issues, such as frequency-domain formulations, multivariate conditioning, or source-space analysis, these approaches often rely on additional assumptions or preprocessing steps that limit their generality (Guo et al., 2010). As such there has been growing interest in causal inference methods that are model-free and more robust to the intrinsic properties of neurophysiological time series, TE being a strong candidate (Vicente et al., 2010).

## 2.4 Transfer Entropy

Transfer Entropy was introduced as an information-theoretic alternative to Granger causality that does not rely on linear models or Gaussian assumptions. TE is a model-free measurement that quantifies directed information transfer by measuring how much the past of one process reduces uncertainty about the future of another, beyond the target's own history. This formulation allows TE to capture nonlinear, asymmetric dependencies that are common in neural systems. Formally, Transfer Entropy from a source process $X$to a target process $Y$is defined as the conditional mutual information between the future of $Y$and the past of $X$, conditioned on the past of $Y$. For discrete-time processes, this is given by

$$T_{X\to Y}=\sum p(y_{t+1}, y_t^{(k)}, x_t^{(l)}) log \frac{p(y_{t+1}|y_t^{(k)}, x_t^{(l)})}{p(y_{t+1}|y_t^{(k)})},$$

where $y_t^{(k)}$and $x_t^{(l)}$denote embedding vectors of past states of $Y$and $X$, respectively (Schreiber, T. 2000). This formulation explicitly quantifies the additional predictive information provided by the source process beyond the target's own history, capturing directed and asymmetric information transfer without assuming linearity or a Gaussian distribution.

By operating directly on probability distributions rather than parametric models, TE provides a principled framework for detecting directed interactions in complex systems (Abdul Razak & Jensen, 2014) and nonstationary data (Vicente et al., 2010). This has led to its increasing adoption in neuroscience, particularly in EEG and MEG studies where linear assumptions are difficult to justify. Importantly, the causal interpretation of TE remains grounded in predictive information flow rather than physical intervention, aligning with the a posteriori nature of neuroimaging data (Wibral et al., 2014).

Practical estimation, despite the advantages of TE, presents significant challenges. Reliable estimation requires accurate inference of high-dimensional conditional probability distributions, making TE sensitive to embedding dimension, finite sample size, and estimator bias. As dimensionality increases, the number of samples required for reliable estimation grows exponentially, motivating representations that reduce state-space complexity without discarding the predictive structure (Papana et al., 2020). These difficulties are particularly acute in EEG and MEG, motivating alternative formulations of TE that improve robustness and tractability.

## 2.5 Learning Temporal Dependencies with Neural Networks

Recent advances in deep learning have introduced new possibilities for causal inference in time-series data (Fang et al., 2018). Neural estimators based on variational bounds and divergence measures have been proposed to approximate mutual information, directed information, and Transfer Entropy directly from data bypassing explicit density estimation (Molavipour et al., 2021). These approaches demonstrate that complex statistical dependencies in principle, can be learned by flexible nonlinear models. Recurrent neural networks, including Long Short-Term Memory (LSTM) networks and Gated Recurrent Units (GRUs), are particularly well suited to this task due to their ability to model temporal dependencies and maintain memory over time (Lindemann et al., 2021; Ubal et al., 2023). By processing ordered sequences, such architectures implicitly approximate the conditional distributions required for causal inference, making them natural candidates for learning directed relationships in time series.

Neural causal discovery frameworks further extend this idea by attempting to infer causal structure directly from data using learnable models. Examples include neural Granger causality approaches (Lindemann et al., 2021) and continuous-time causal discovery methods such as DYNOTEARS (Zhu et al., 2021). While these methods highlight the potential of learning-based approaches they do often operate based on raw amplitude signals and may require large datasets and careful optimisation to achieve stable performance. As a result representation remains a central challenge in learning-based causal inference, where the choice of input representation strongly influences model stability, interpretability, and data efficiency, particularly in high-noise neuroimaging contexts (Deng et al., 2025). This motivates the exploration of alternative representations that reduce estimation complexity while preserving essential temporal structure.

## 2.6 Implications for Causal Inference

Statistical dependencies such as consistent time-lagged dependencies and physical field dynamics of neurons render information transfer directionality inherently fragile in EEG and MEG data (Murakami & Okada, 2006) with noise and source mixing from sensor arrays further obscuring temporal asymmetries (Larson & Taulu, 2018). As a result, causal estimates may vary substantially across time windows or trials and depend sensitively on analysis parameters such as the lag conditions needed for information transfer calculations. In addition, unobserved common drivers, and indirect interactions further confound interpretation. Apparent directional relationships may arise from shared inputs or mediated pathways rather than direct causal influence. This is particularly true in high-dimensional neural systems where not all relevant sources are observable (Reid et al., 2019).

These factors imply that information transfer directionality should not be interpreted as a stable property of individual signal segments. Instead, reliable inference often critically requires representations. However, aggregation strategies that mitigate noise, nonstationary, and mixing effects might also be of crucial importance (David et al., 2006).

## 2.7 Practical Challenges in TE Estimation

Classical estimators of Transfer Entropy typically rely on estimating conditional mutual information from continuous-valued time series using kernel density estimation or k-nearest-neighbour methods (Vicente & Wibral, 2014). While these approaches are theoretically well founded, they suffer from well-known bias–variance trade-offs and scale poorly with increasing embedding dimension. As the dimensionality of the joint state space grows, the number of samples required to obtain reliable probability estimates increases exponentially, making direct TE estimation impractical in many real-world settings (Zhu et al., 2015). This challenge is particularly acute for EEG and MEG data. When raw amplitude values are used directly, the estimator must resolve fine-grained variations in signal magnitude across multiple time lags, resulting in sparse sampling of the underlying probability space. In such cases, estimation error is dominated not by model mismatch but by insufficient data coverage, leading to unstable or biased TE estimates (Shorten et al., 2020).

When raw amplitude values are used directly for Transfer Entropy estimation, the estimator must resolve fine-grained variations in signal magnitude across multiple time lags, resulting in sparse sampling of a high-dimensional probability space. In EEG and MEG, where recordings are noisy, nonstationary, and constrained by limited trial counts, estimation error is therefore often dominated by insufficient data coverage rather than model mismatch, leading to unstable or biased estimates (Wollstadt et al., 2014). Consequently, the difficulty of TE estimation depends not only on embedding dimensionality but on the informativeness of the chosen representation: raw amplitudes place a high burden on the estimator by requiring predictive dependencies to be inferred from unstructured continuous states, while overly coarse representations reduce variance at the cost of discriminative power (Pinzuti et al., 2020). From an information-theoretic perspective, effective representations concentrate predictive information into distinguishable states while suppressing noise and redundancy (Shwartz Ziv & LeCun, 2024).

More generally, the complexity of TE estimation depends not only on dimensionality but on the *informativeness of the representation*. Representations that encode multiple complementary features, such as relative magnitude, local variation, or structural configuration, can provide richer syntactic cues that help differentiate signal states without requiring finer discretisation or larger datasets (Faes et al., 2016). From this perspective, increasing representational structure can reduce estimation burden by distributing information across multiple interpretable dimensions, rather than relying solely on raw amplitude or minimal symbolic codes.

This insight motivates the exploration of symbolic representations that preserve richer temporal and morphological structure while remaining computationally tractable. A representation then that can encode multiple complementary aspects of local signal structure, such as relative magnitude, variation, or configuration, could increase state separability without requiring finer discretisation or substantially larger datasets, thereby reducing estimation burden through representational efficiency rather than increased model complexity.

## 2.8 Symbolic and Approximate Transfer Entropy

Symbolic formulations of Transfer Entropy discretise continuous signals into symbolic sequences prior to entropy estimation, reducing estimator variance and improving robustness to noise. Approaches such as Symbolic Transfer Entropy (Dimitriadis et al., 2016) and ordinal or phase-based methods (Zhang, Lin, & Shang., 2017) simplify the state space but often rely on limited symbolic alphabets, constraining expressiveness. In symbolic formulations of Transfer Entropy, the continuous-valued time series $X_t$ and $Y_t$ are first mapped to discrete symbolic sequences $S_t^X$ and $S_t^Y$ via a symbolisation function, in this case the geometric embedding of polygon into the time series. Transfer Entropy is then computed over the resulting symbolic processes as:

$$T_{X\to Y}^{sym} = \sum p(s_{t+1}^{Y}, s_t^{Y,(k)}, s_t^{X,(l)}) log \frac{p(s_{t+1}^{Y} | s_t^{Y,(k)}, s_t^{X,(l)})}{p(s_{t+1}^{Y} | s_t^{Y,(k)})},$$

where $s_t^{Y,(k)}$ and $s_t^{X,(l)}$ denote symbolic embedding vectors (Dimitriadis et al., 2016). This formulation preserves the directional and conditional structure of Transfer Entropy while reducing the effective state space, improving robustness to noise and finite-sample effects at the cost of representational granularity.

While symbolic representations discard some fine-grained signal detail, they capture transition statistics and temporal regularities closely related to the foundations of TE. From a statistical perspective, symbolic entropy measures can be interpreted as proxies for directional predictability under mild assumptions (Porfiri & Ruiz Marín, 2019). Geometric symbolic representations further extend this idea by encoding morphological structure rather than raw amplitude. Such abstraction emphasises recurring temporal motifs and shape transitions, offering improved robustness to noise and nonlinearity (Dimitriadis et al., 2016). However, existing symbolic methods often trade computational tractability for representational impoverishment, motivating the development of richer symbolic encodings.

## 2.9 Extending Symbolic Representations Using Shape

A range of symbolic representations have been proposed to reduce the dimensionality of continuous time series while preserving information relevant for learning and inference. One widely used approach is Symbolic Aggregate approXimation (SAX), which partitions a signal into fixed-length windows and represents each segment by its mean amplitude, subsequently discretised into symbols. While SAX is computationally efficient and robust to noise, its reliance on averaging leads to a loss of local temporal structure and shape information, limiting its sensitivity to fine-grained dynamics (Lin et al., 2003). ABBA (Aggregation by Breakpoint Approximation), an extension of SAX that uses adaptive piecewise aggregation to capture local signal geometry more effectively. They emphasize that classical SAX's mean-based aggregation produces overly coarse representations, motivating geometric and adaptive symbolization schemes that retain shape transitions and temporal motifs (Chen & Güttel, 2023).

Ordinal pattern methods such as those in Amigó (2010), in *Permutation Complexity in Dynamical Systems* provides the foundational formulation and encode time series based on the relative ordering of values within short temporal windows, discarding absolute amplitude information while preserving local rank structure. Such approaches offer robustness to noise and monotonic transformations but collapse distinct signal segments that share the same ordering into identical symbols, which would substantially limit representational expressiveness of the time series data (Amigó, 2010). Polygonal decomposition provides a structured alternative to conventional symbolic encodings by representing a continuous time series as a sequence of geometric primitives derived from local signal structure. Unlike shapelet-based methods, the polygonal primitives used in the present work are not selected or optimised for discriminative performance but arise deterministically from local signal structure and serve as symbolic states within a task-independent representation.

Extrema-based segmentation instead defines segments by identifying local maxima and minima using turning points to delimit geometric primitives. This approach is inherently event-driven because each polygon corresponds to a meaningful change in signal direction and captures local dynamics such as rise–fall structure, asymmetry, and curvature, as a result, geometric decomposition into polygons adapts naturally to variations in signal frequency and amplitude, producing a representation that is sensitive to intrinsic temporal organisation rather than externally imposed windowing or supervised learning. From a neurophysiological perspective, extrema-based shape representations are well aligned with event-based models of neural processing. Neural systems often encode information through transient events such as bursting or plateau potential, state transitions, and changes in activity rather than sustained absolute levels all of which exhibit geometric identities (Patrascu, 2025) both morphologically (Debanne et al., 2011) and harmonically (Nowak et al., 2003).

## 2.10 Shape-Based Symbolic Dynamics

In EPSTE, each segment of a time series is represented not as a scalar value or low-dimensional symbol, but as a geometric object encoding multiple aspects of local temporal structure. Treating triangles as symbolic states allows temporal dynamics to be modelled as transitions between geometric configurations. Rather than estimating dependencies directly from continuous amplitudes, the system is described by a sequence of shape states whose transitions reflect underlying signal dynamics. This reframes the time series as a symbolic dynamical system, where causal influence can be expressed in terms of how the state sequence of one signal constrains or predicts transitions in another. Within this framework, entropy is defined over distributions of shape states or shape transitions, rather than over raw signal values.

In EPSTE's simplest case, successive extrema define triangular primitives ($\chi$-1, $\chi$, $\chi$+1), where each triangle captures relative amplitude on the central $\chi$ value representing **ordinal position in space**, with area of the triangular primitive to represent **magnitude of change** over the triplet, and apical/basal angle representing a **directional correlate of change** over the triplet. These geometric primitives form a discrete but structured state space in which each state corresponds to a particular local morphological pattern rather than a pointwise signal value. The entropy of a shape sequence reflects the diversity and predictability of local morphological patterns, while conditional entropies capture how knowledge of one sequence reduces uncertainty about another. Transition entropy between shape states therefore provides a natural basis for approximating Transfer Entropy, as it directly operationalises the core notion of directional predictability in a reduced, structured state space.

Crucially, geometric primitives encode multiple complementary features—such as local variation, and geo-temporal configuration—within a single symbolic unit. This increases state separability without requiring finer discretisation or higher embedding dimension, improving estimation efficiency in data-limited settings. By concentrating predictive information into structured symbolic transitions, shape-based symbolic dynamics offer a principled compromise between representational richness and computational tractability, forming a suitable foundation for learned approximations of Transfer Entropy. Approaching time series as sequences of structured symbolic motifs rather than raw amplitude fluctuations, reflects a broader tradition in scientific modelling in which understanding arises through appropriate representational languages. Similarly, raw neural time series contain fine-grained amplitude information that is both noisy and high-dimensional, making direct estimation of causal relationships data-intensive and unstable. Polygon representations can be viewed as an intermediate language in which temporal structure is expressed through discrete, information-bearing units rather than raw signal values. In this view, structured symbolic elements function analogously to syllables in speech: they compress continuous signals into syntactically meaningful primitives that preserve relational information while reducing representational burden. Such representations facilitate learning and estimation by exposing regularities in temporal organisation, rather than requiring inference directly from unstructured amplitude fluctuations.

# 3. Methodology

This project develops a learning-based framework for estimating directed information flow between time-series signals and motivated by the limitations of classical Granger Causality and conventional Transfer Entropy (TE) estimators when applied to nonlinear and nonstationary neural data, in this case, MEG (Barnett et al., 2009). Rather than estimating TE directly from raw amplitude time series, the proposed pipeline introduces a geometric–symbolic representation, termed Embedded Polygonal Symbolic Transfer entropy (EPSTE), of temporal dynamics and trains a recurrent neural network (RNN) to learn the mapping, given by matrices of transition states, between these representations and reference TE values. Specifically, a Gated Recurrent Unit (GRU) architecture is employed to model local temporal dependencies within sliding windows of the transformed sequences, enabling the network to capture transient, direction-specific dynamics rather than relying solely on global averages.

## 3.1 Data

Data were provided by the Sussex Centre for Consciousness Science and consist of secondary MEG recordings from a psychedelic study. Five participants were used in training of the RNN. Because the work uses pre-existing, anonymised secondary data, no additional data acquisition or participant recruitment was performed for this work.

For each subject, approximately 10 minutes of source-reconstructed MEG data were available, originally sampled at 600 Hz. Data were provided segmented into 2-second epochs, which conveniently aligns with the bag/trial structure used later in the multi-instance learning setup. Artefact epochs (e.g., eye blinks, muscle activity) had already been removed during upstream preprocessing. This meant that the number of usable epochs varied by subject (approximately 100–200 epochs per subject. Signals were source-reconstructed and parcellated using the AAL atlas (90 regions), yielding an array of 90 regional time series per epoch. The dataset had also been high-pass filtered at 1 Hz upstream to suppress slow drifts.

To standardise signal quality and reduce nuisance spectral components prior to symbolic feature construction and modelling, each epoch underwent a minimal preprocessing pipeline implemented in Python (SciPy): Notch filtering at 50 Hz and 100 Hz to suppress UK mains interference and its first harmonic. Low-pass filtering (default 45 Hz) to attenuate higher-frequency components that are less relevant for the present methodological analysis and are more likely to contain residual noise. Down sampling from 600 Hz to a target sampling rate of 250 Hz using polyphase resampling, reducing computational load while retaining temporal resolution well above the post-filter bandwidth. Channel-wise z-scoring (mean removal and variance normalisation per parcel) was used to place regions on a comparable scale and prevent any single parcel's absolute amplitude from dominating downstream feature extraction. All filtering operations were performed using the zero-phase forward–backward filtering (filtfilt) library to avoid introducing phase delays that could distort temporal relationships relevant to directed-dependence estimation.

MEG recordings were loaded from MATLAB files and decomposed into individual trials. For each file, the primary signal array was automatically identified and reshaped to a consistent channel / time / trials format, after which each trial was independently pre-processed using the pipeline described above. Trial-level metadata, including subject identity, drug condition, and trial index, were retained to support hierarchical aggregation in subsequent analyses. To prevent subject-level information leakage, data were split into training, validation, and test sets at the subject level. All trials and derived windows from a given subject were assigned exclusively to a single split. The number of MEG channels used in the study equated to 16, which equates to 8 general regions each with a left and right hemispherical split.

The programming elements for this experiment were executed using the Python 3.11.4 scripting language and utilised the following Python libraries: numpy, Matplotlib, math (atan, degrees, gcd), pandas, torch (torch.utils.data: DataLoader, TensorDataset, Dataset), collections (Counter, defaultdict), itertools (permutations), random, copy, scipy.stats (pearsonr, wilcoxon, spearmanr, ttest_rel), scipy.io (loadmat), scipy.signal (butter, filtfilt, detrend, resample_poly, iirnotch), mne, re, os, and pathlib (Path).

## 3.2 Geometric Decomposition

Each time series was decomposed into a sequence of local geometric primitives computed from overlapping triplets of consecutive samples. For each valid time index $t$, the triplet $(x_{t-1}, x_t, x_{t+1})$defines a local triangular primitive (as can be observed in figure 1) from which three continuous morphology features were extracted: (i) triangle area, (ii) an angle/orientation term derived from the relative slopes of adjacent segments, and (iii) an amplitude-change term. This produces three feature sequences of length $T$-2for a time series of length $T$, yielding a local geometric representation of waveform shape.

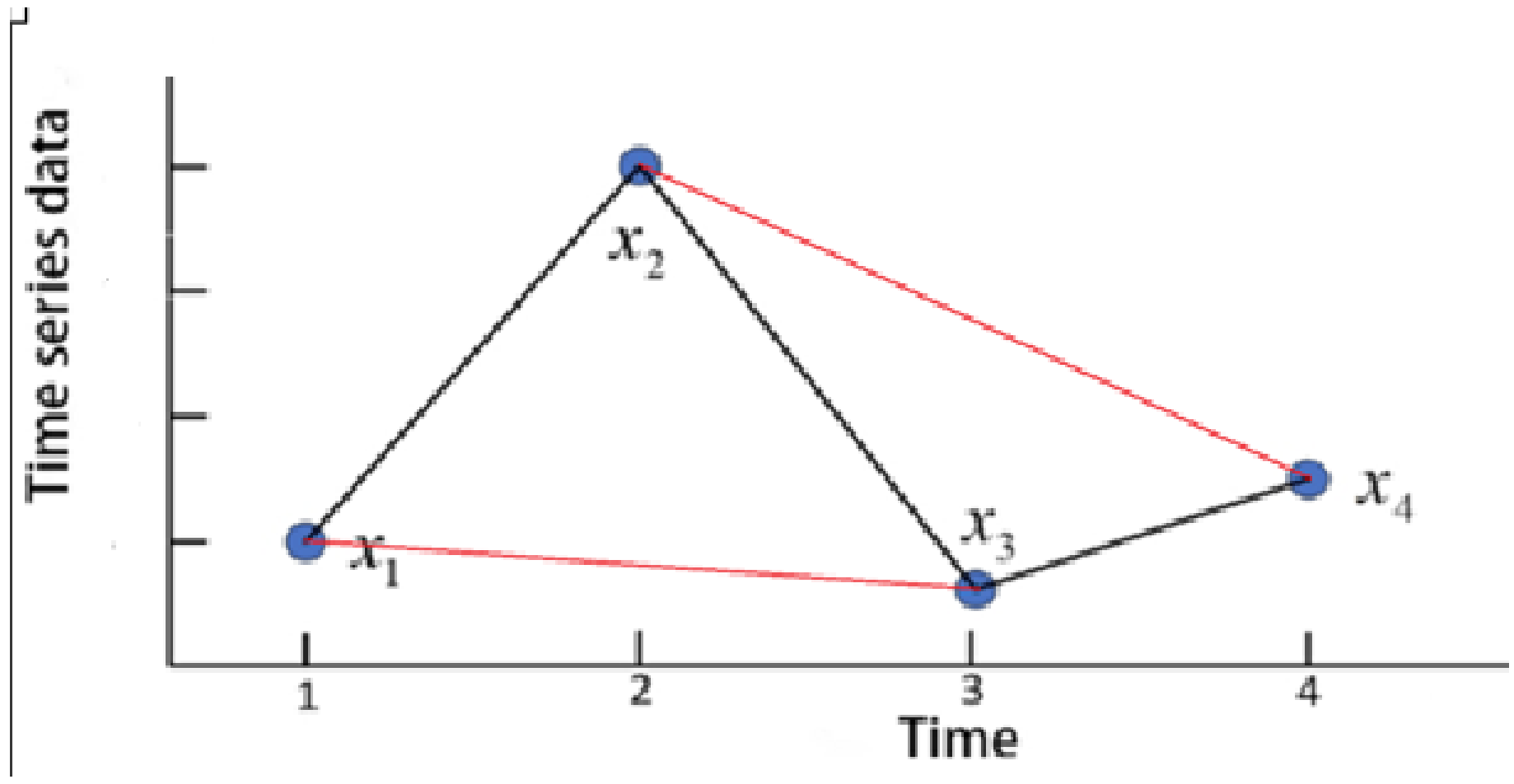

*Figure 1. Schematic illustration of geometric time-series decomposition.*
*Consecutive samples of a univariate time series are embedded in Cartesian space and connected to form local geometric primitives. Triplets of adjacent points define polygonal elements (e.g., triangles) whose geometric properties—such as area, internal angles, and relative orientation—characterise local morphological structure in the signal.*

To convert these continuous features into discrete symbols suitable for symbolic TE estimation and neural network inputs, each feature dimension was discretised using quantile-based global bins learned from the training set only. At each time index, the three discretised bin indices (area-bin, amplitude-bin, angle-bin) were then packed into a single categorical token, yielding a symbolic sequence that preserves local morphological transitions while reducing sensitivity to absolute amplitude scaling and measurement noise. More complex geometric primitives were avoided to limit representational complexity and reduce sensitivity to noise, while still retaining sufficient local structure to support directed dependency estimation.

## 3.3 Symbol Vocabulary and Feature-Library Construction

The discretisation scheme implicitly defines a finite library of possible triangle categories: the Cartesian product of area, amplitude, and angle bins. Given $n_A$bins for area, $n_M$bins for amplitude, and $n_N$bins for angle, the maximum symbol vocabulary size is $n_A \times n_M \times n_N$. Each observed triangle primitive maps to exactly one token in this library via binning and index packing. In practice, not all possible tokens are necessarily realised in the data; therefore, the effective vocabulary was constructed from the set of tokens observed in the training set, with an optional "UNK" token for unseen symbols, and the resulting mapping was reused unchanged for the validation and test data.

Binning strategy is treated as an integral part of the symbolisation methodology rather than a mutable hyperparameter. Because the triangle-based EPSTE representation and the raw-amplitude baseline inhabit different feature spaces, their empirical symbol distributions differ by construction and enforcing identical bin edges or symbol frequencies would therefore impose artificial constraints and introduce representational bias. However, it should be noted the specific discretization values were arbitrarily imposed and not learnt constituting a fundamental limitation. Equivalence between conditions are ensured procedurally by applying the same quantile-based binning rule rather than by matching resulting symbol statistics. Any differences in symbol distributions thus reflect genuine representational structure rather than optimisation method or hyperparameter.

## 3.4 Network Label Generation

Target labels were generated as trial-level symbolic Transfer Entropy (TE) values computed for each MEG channel pair. For a given channel pair $(x, y)$, the pipeline first transformed each continuous time series into a symbolic sequence and then estimated directional $STE_{x\rightarrow y}$. These trial-level TE values served as supervision targets during model training, with aggregation across trials used only for evaluation and connectivity analysis.

Symbolisation proceeded as follows. For each time series, local geometric features were extracted using overlapping triplets of samples, yielding three continuous features per valid time index: triangle area, an implied orientation/angle term, and amplitude change. To ensure consistent discretisation across all samples, global quantile bin edges were estimated using training data only. A random subsample (up to max_pairs) of training channel pairs was used to collect feature distributions, and quantile-based bin edges were computed separately for each feature dimension. These bin edges were then fixed and reused unchanged for train, validation, and test data. Each feature value was discretised according to the pre-fit bins, and the resulting feature bins were combined into a single categorical symbol per time index, producing a symbolic sequence for $x$ and a symbolic sequence for $y$. Directional STE was computed using a joint-count estimator. For a given lag $l$, TE from $x$ to $y$ was estimated from joint distributions over symbolic tuples of the form:

$$(y_{t+1}, y_t, x_{t-l}),$$

with marginal terms obtained by summation. TE was reconstructed via the standard conditional entropy formulation (Schreiber, T. 2000):

$$TE_{x\to y}=H(y_{t+1}|y_t)-H(y_{t+1}|y_t, x_{t-l}),$$

The implementation used entropy differences computed from joint counts, with an optional Miller–Madow correction applied to reduce small-sample bias. Estimates were clamped to be non-negative.

TE was evaluated over a small, physiologically motivated lag set (e.g. $l \in \{10,20,30\}$), and the maximum TE across lags was retained as the final trial-level target value. These lags correspond to shifts in the discretised symbolic sequence rather than raw signal samples and therefore represent local temporal dependencies rather than precise physiological delays. Restricting the lag range during supervision improves statistical stability of the TE estimates and avoids diluting temporally specific interactions through averaging and also because STE estimation becomes noisy very quickly as lag increases (Faes et al., 2014). Larger lag ranges were explored subsequently during aggregation and surrogate analyses, where physiological interpretation was the primary focus. The maximum TE across the evaluated lag set was selected as the final target value, reflecting the assumption that directed interactions may occur at a characteristic delay rather than being uniformly distributed across lags. Averaging across lags would dilute temporally specific effects and bias estimates toward zero.

## 3.5 Neural Network Architecture and Parameters

To predict trial-level directed Transfer Entropy (TE) values from symbolic representations of MEG channel pairs, a neural architecture was designed to operate on bags of symbolic temporal windows, rather than on individual time points. This reflects the fact that TE is defined over extended temporal structure and may be expressed only intermittently across a trial. The resulting model adopts a multiple-instance learning (MIL) formulation, in which each trial-level channel pair is represented as a set of local windows, and the network learns to aggregate information across windows to produce a single scalar TE estimate.

## 3.6 Input Representation

Each input sample corresponds to a single trial and channel pair $(x, y)$, represented as a bag of overlapping temporal windows extracted from the symbolic sequences of the two channels. For each window $w$, two discrete symbolic sequences are provided: a source sequence $s_x^{(w)}$and a target sequence $s_y^{(w)}$. Symbols are integer-valued categorical tokens derived from the triangle-based geometric discretisation described in the previous section. Symbolic sequences were segmented into overlapping temporal windows of fixed length $W$, with stride equal to the window length to avoid redundant overlap. A structural offset equal to the maximum evaluated lag was applied to ensure that source symbols $x_{t-l}$and target symbols $y_{t+1}$were temporally well-defined within each window and to prevent information leakage across causal boundaries.

Because source and target channels play asymmetric roles in directed information transfer, separate vocabularies and embedding layers are used for $x$and $y$. Discrete symbols from the source and target sequences are mapped to continuous vectors using learned embedding layers:

$$e_x: S_x \rightarrow \mathbb{R}^d, e_y: S_y \rightarrow \mathbb{R}^d,$$

where $d$ is the embedding dimension. Embeddings for the two channels are not shared, allowing the model to learn asymmetric representations appropriate for causal source and target roles. At each timestep, the embedded source and target symbols are concatenated to form a joint representation that is passed to the temporal encoder. Local temporal structure within each window is modelled using a gated recurrent unit (GRU). For a given window, the concatenated embeddings are processed sequentially by a GRU with hidden dimensionality $H$, producing a sequence of hidden states. A fixed-length window representation is obtained from the GRU output (e.g. using the final hidden state).

GRUs were selected due to their computational efficiency and stability when modelling moderately sized time series, as well as their ability to capture short-range temporal dependencies without the parameter overhead of deeper recurrent architectures.

## 3.7 Multiple-Instance Attention Pooling

Since not all windows are expected to contribute equally to directed information flow, window-level representations are aggregated using an attention-based MIL pooling mechanism. For each window representation $h_w$, a scalar attention weight $\alpha_w$is computed via a learnable scoring function:

$$\alpha_w = softmax(v^\top tanh(W h_w)),$$

where $W$ and $v$ are trainable parameters. The trial-level representation is then formed as a weighted sum of window embeddings:

$$H = \sum_w \alpha_w \, h_w.$$

This mechanism allows the model to focus on temporally localised segments that are most informative for causal influence, rather than assuming uniform contribution across time. Critically, attention weights were used solely as a mechanism for adaptive aggregation and were not interpreted as direct measures of causal importance in this work.

## 3.8 Output Layer and Training Objective

The aggregated trial-level representation is mapped to a scalar prediction using a linear output layer. The network is trained to minimise mean squared error (MSE) between predicted and ground-truth TE values. The model output was constructed as a regressor estimation rather then a classification of STE.

## 3.9 Hyperparameters and Optimisation

Key architectural hyperparameters include the embedding dimension $d$, GRU hidden size $H$, number of recurrent layers (set to one), attention projection dimension, and dropout rate. These values were chosen to balance representational capacity with regularisation, given the size and structure of the dataset. A modest dropout rate was applied to reduce overfitting without suppressing informative temporal structure.

Training was performed using the Adam optimiser with a small learning rate of $1\times10^{-4}$, and early stopping based on validation loss was employed to prevent over-training. To stabilise optimisation, trial-level TE targets were standardised using the mean and variance computed from the training set only. Predictions were transformed back to the original TE scale for evaluation and reporting. Importantly, all architectural and optimisation parameters were held constant across model variants, including baseline and triangle-based representations, to ensure that observed performance differences were attributable to the input representation rather than differences in model capacity.

## 3.10 Model–Target Alignment

The architecture is explicitly aligned with the semantics of the learning target. Ground-truth TE values are defined at the trial level and reflect aggregated directional dependencies over time. By operating on bags of local windows and using attention-based aggregation, the network mirrors this structure, enabling interpretable and stable estimation of directed information flow from symbolic temporal data.

## 3.11 Model Predictions and Aggregation

Model predictions are generated at the window level, capturing local dynamics, and are subsequently aggregated hierarchically to the trial (bag) level and finally to the pair (edge) level. This aggregation strategy is shown to be essential for stabilising causal estimates and recovering consistent directed structure across noisy local predictions.

## 3.12 Control Condition: Standard Symbolic Transfer Entropy Baseline

To evaluate whether improvements in TE prediction were attributable to the proposed geometric symbolic representation rather than model capacity or training procedure, a control condition based on standard symbolic Transfer Entropy (TE) was implemented.

In this baseline, ground-truth TE labels were generated directly from discretised raw MEG time series, without geometric or triangle-based feature extraction. Each channel time series was discretised into categorical symbols using global quantile bins learned from the training set only. These symbols were then used to compute directional symbolic TE using the same joint-count estimator and lag set as in the triangle-based condition. Specifically, for each trial-level channel pair $(x, y)$, symbolic $TE_{x \to y}$was computed across an identical lag range, and the maximum TE across lags was retained as the target value. All estimator settings, including optional Miller–Madow bias correction and non-negativity clamping, were held constant between conditions. Crucially, the neural network architecture, hyperparameters, optimisation procedure, and data splits were identical between the baseline and triangle-based models. The same attention-based multiple-instance learning architecture was used to predict trial-level TE values from bags of symbolic windows, differing only in the input symbol sequences and corresponding target labels. This control condition therefore isolates the contribution of the geometric symbolic representation itself. Any observed differences in predictive accuracy, stability, or aggregation behaviour can be attributed to differences in representational structure, rather than estimator choice or model capacity.

## 3.13 Surrogate Testing

To verify that estimated Transfer Entropy (TE) values reflected genuine directed temporal structure rather than spurious correlations, phase-randomisation surrogate testing was performed on a subset of 200 held-out channel pairs. In this procedure, the source time series was repeatedly phase-randomised, preserving its power spectrum while destroying temporal dependencies, and symbolic TE was recomputed against the unchanged target. Observed TE values were compared to the resulting null distributions to obtain empirical p-values and z-scores, defined as

$$z = \frac{TE_{\mathrm{obs}} - \mu_{\mathrm{null}}}{\sigma_{\mathrm{null}}}$$

Surrogate testing served as a statistical validation step, ensuring that both ground-truth TE estimates and model predictions exceeded chance levels. Importantly, surrogate data were not used during model training and played no role in optimisation.

## 3.14 Evaluation Metrics and Statistical Analysis

Model performance was evaluated using a combination of error-based, correlational, and distributional metrics, chosen to reflect both predictive accuracy and stability of estimated directed information flow. Because Transfer Entropy (TE) estimates can exhibit heavy-tailed and non-Gaussian distributions (García-Medina & González Farías, 2020), evaluation emphasised robust non-parametric statistics rather than mean-based measures alone. Prediction accuracy was quantified using mean squared error (MSE) and Pearson correlation coefficient (r) between predicted and ground-truth TE values. These metrics were computed at multiple levels of aggregation, including window-level predictions, trial-level (bag-level) predictions, and aggregated pair-level estimates. This multi-scale evaluation allows assessment of both local prediction fidelity and the recovery of stable directed dependencies across trials.

To assess robustness, median error, interquartile range (IQR), and distributional spread of predictions were examined, particularly at the pair level where TE is typically interpreted as a connectivity measure. Variance reduction under aggregation was treated as an important indicator of estimator stability rather than overfitting. For direct comparison between representational conditions, paired non-parametric statistical testing was employed. Specifically, Wilcoxon signed-rank tests were used to compare error distributions between the triangle-based representation and the standard symbolic baseline across matched channel pairs. This test was chosen due to its robustness to non-normality and outliers, and because comparisons were performed on paired observations derived from identical data splits and model architectures.

# 4. Analysis and Results

## 4.1 Experimental Setup

The experimental set up is framed as a comparison of multiscale causal inference when local geometric embedding dimension of the time series is increased to produce motifs of polygons. It should be noted this experiment is designed not just to compare conventional test-train predictive accuracy of the time series representation methods but to determine whether stable directed dependencies can be recovered from local observations through aggregation towards a central limit. Accordingly, success is assessed not by window-level accuracy alone, but by whether aggregation yields stable, interpretable, and statistically viable pair-level causal estimates in which causal structure is expected to emerge only after aggregation across noisy local observations. Within this experimental frame any weak and unstable window-level performance is not interpreted as methodological breakdown but an expected consequence of estimating distributed causal structure from locally noisy nonlinear time series data.

In order to produce an empirically controlled experiment two causal representations were compared. The stated EPSTE method, and a raw amplitude signal discretised baseline representation of the time series data. To ensure statistical credibility both time series representation was trained using an identical multiple instance learning (MIL) attention architecture. Identical ground truth supervision of the models was derived from transfer entropy estimates, computed under constrained lag windows approximately consistent with realistic neural propagation times. Inference was explicitly evaluated at three hierarchical scales: window-level (local dynamics), bag-level (trial-wise aggregation), and pair-level (directed edge estimates). Performance metrics were selected to reflect both accuracy and structural fidelity, including MAE, RMSE, Pearson's r, Spearman's ρ, and $R^2$. Rather than a single aggregate loss and statistical comparison, representations were conducted using paired, non-parametric tests. This ensured robustness to non-Gaussian error distributions and preserving pairwise correspondence.

## 4.2 Window-Level Results

At the window level, both representations exhibited high variance and weak correspondence with ground-truth Transfer Entropy (TE) values. On the held-out test subject (Subject 5), the EPSTE (triangle-based) representation achieved a Pearson correlation of $r = 0.48$, while the raw-amplitude baseline achieved a comparable correlation of $r = 0.52$. Spearman rank correlations were similarly modest ($\rho \approx 0.40$ for both methods), indicating limited monotonic association between window-level predictions and true TE values. Despite these moderate correlation coefficients, absolute error remained high for both approaches. The EPSTE model exhibited a window-level RMSE of 1.36 and MAE of 1.10, while the baseline achieved substantially lower absolute error (RMSE = 0.05, MAE = 0.04). However, coefficients of determination were poor or negative for both methods (EPSTE $R^2 = -0.94$; baseline $R^2 = 0.10$), indicating that window-level predictions failed to reliably explain variance in the underlying STE labels.

## 4.3 Bag-Level Results (Trial Aggregation)

Aggregating window-level predictions at the trial (bag) level resulted in a substantial improvement in performance for both representations, confirming that directed information cannot be reliably inferred from short local segments alone but instead emerges through integration across time. This behaviour is consistent with the multiscale inference framework adopted in this work, in which causal structure is expected to arise only after aggregation of noisy, local observations. For the held-out test subject (Subject 5), the EPSTE model achieved a strong correspondence with ground-truth Transfer Entropy values at the bag level (figure 2), with a Pearson correlation of $r = 0.86$ and a coefficient of determination of $R^2 = 0.74$. Absolute error was substantially reduced relative to the window level (MAE = 0.021), indicating that aggregation enabled the model to recover a large proportion of the variance in the underlying TE estimates.

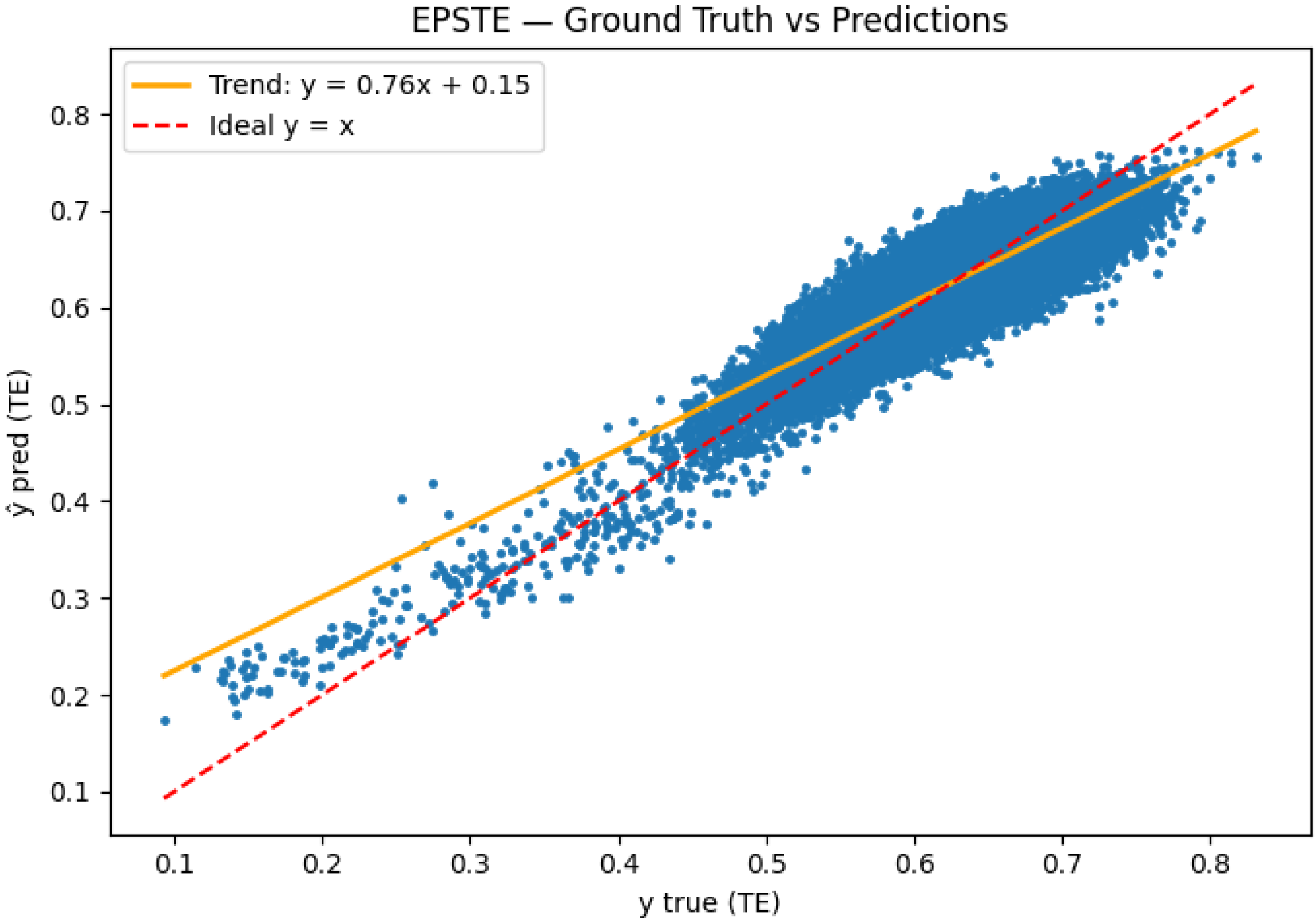

*Figure 2. Scatter plot comparing predicted EPSTE values against ground-truth symbolic transfer entropy at the pair level for the held-out test subject. The dashed red line indicates the ideal identity relationship (y = x), while the solid line shows the best linear fit to the predictions, illustrating strong correspondence with slight compression at higher TE values. The tight clustering around the identity line indicates accurate recovery of directed information strength under aggregation.*

In contrast, the raw-signal baseline achieved a lower Pearson correlation (r = 0.68) and reduced explanatory power ($R^2$ = 0.46), alongside higher absolute error (MAE = 0.031). While aggregation improved baseline performance relative to the window level, the raw representation failed to capture directed dependencies with the same fidelity as EPSTE.

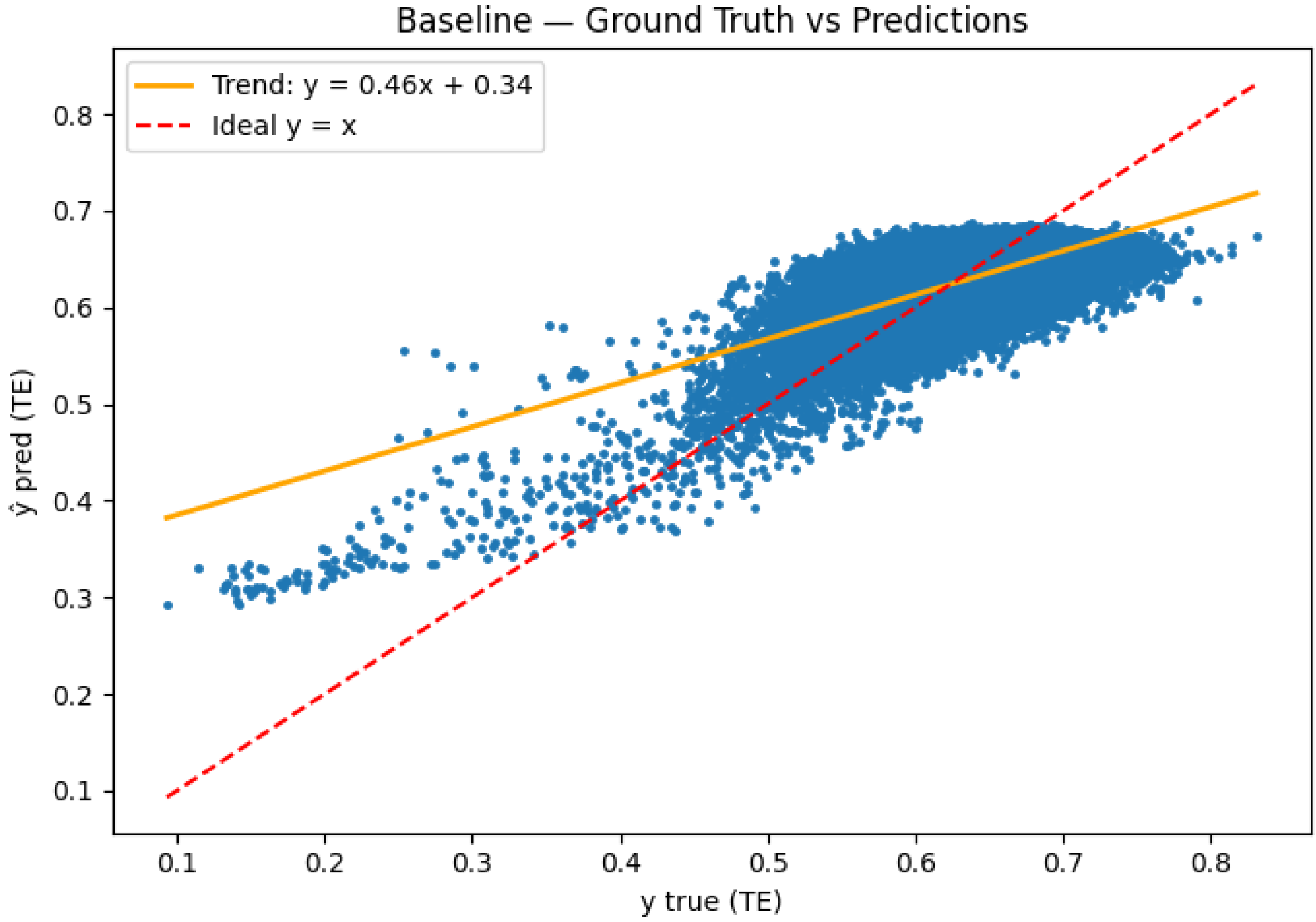

*Figure 3. Scatter plot comparing predicted transfer entropy values from the raw-signal symbolic baseline against ground-truth TE at the pair level for the held-out test subject. The dashed red line denotes the ideal identity relationship (y = x), while the fitted trend line reveals systematic compression toward the mean, with reduced slope and offset bias. Compared to EPSTE, predictions exhibit greater dispersion and attenuated dynamic range, indicating diminished precision in recovering directed information strength despite high overall correlation.*

Qualitative inspection of bag-level scatter plots (Figures 2 and 3) supports these findings. EPSTE predictions exhibit tighter clustering around the identity line and reduced dispersion compared to the baseline, suggesting improved structural alignment with the ground-truth TE values.

## 4.4 Pair-Level Results (Primary Outcome)

Pair-level aggregation constitutes the primary outcome of this study, as Transfer Entropy is fundamentally defined over directed interactions between pairs of processes rather than individual windows or trials. At this level of analysis, predictions are aggregated across all trials for each directed channel pair, maximising sample support and substantially reducing variance arising from trial-level noise. This aggregation step is therefore theoretically aligned with both the formal definition of TE and the multiscale inference framework adopted in this work.

For the held-out test subject (Subject 5) pair-level aggregation produced the strongest and most interpretable results for both representations ( figure 4) , with the EPSTE model exhibiting near-perfect correspondence with ground-truth TE estimates. Specifically, EPSTE achieved a Pearson correlation of $r \approx 0.99$ and a Spearman rank correlation of $\rho \approx 0.99$, indicating both high linear agreement and preservation of rank-order structure across directed edges. The coefficient of determination was correspondingly high ($R^2 \approx 0.98–0.99$), demonstrating that the vast majority of variance in the ground-truth TE values was recovered by the model

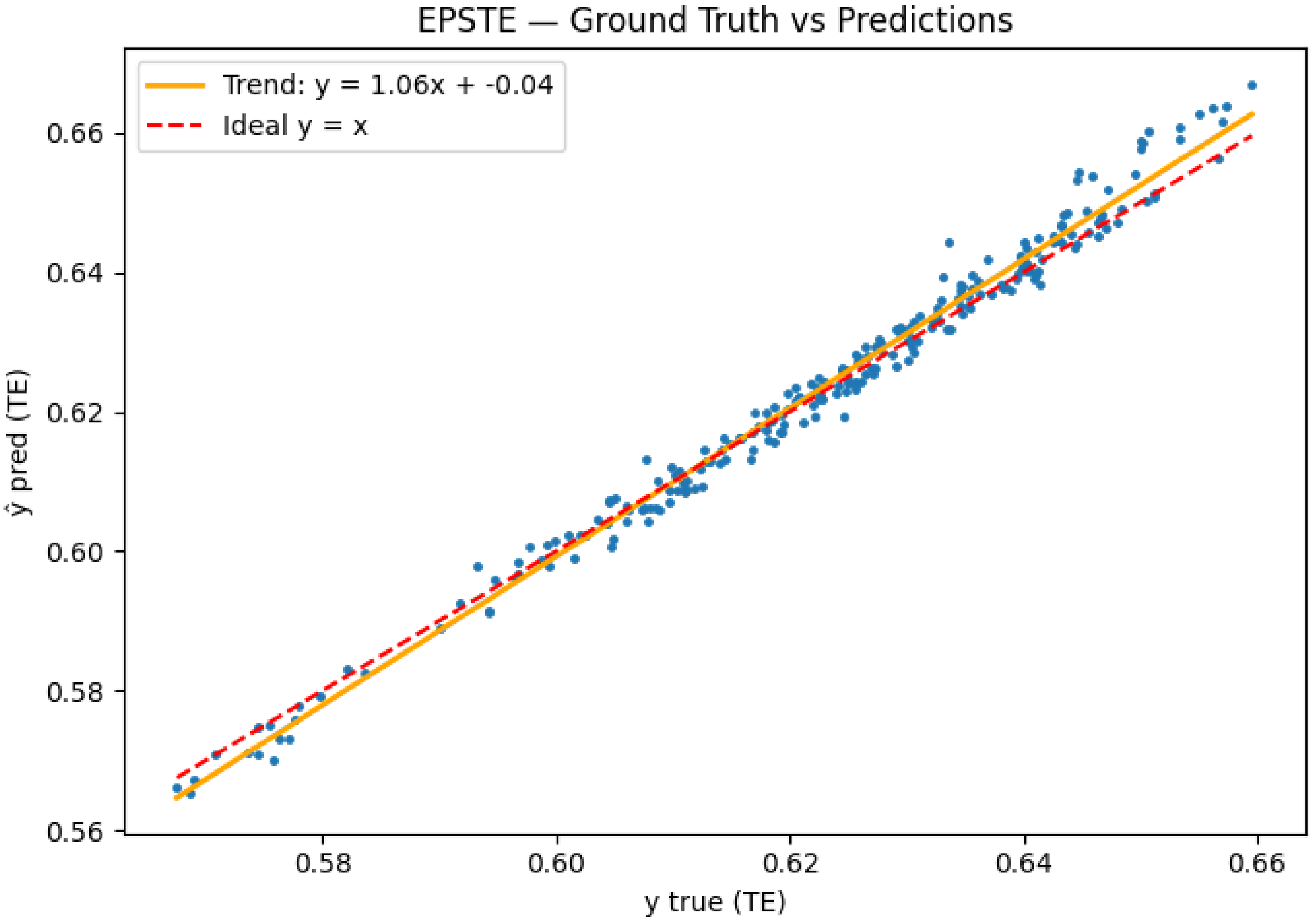

*Figure 4. Pair-level comparison of predicted transfer entropy values from the EPSTE model against ground-truth TE for the held-out test subject. The dashed red line indicates the ideal identity relationship (y = x), while the fitted trend line shows near-unity slope and minimal bias. The tight clustering around the identity line demonstrates high quantitative accuracy and preservation of dynamic range, supporting the notion that geometric symbolic representation supports precise recovery of directed information strength under aggregation.*

Absolute error metrics further support this conclusion. EPSTE achieved a mean absolute error of MAE ≈ 0.0018 and an RMSE ≈ 0.0029 at the pair level, reflecting highly accurate quantitative recovery of TE magnitudes. Pair-level scatter plots show tight clustering around the identity line, confirming that EPSTE preserves both relative ordering and absolute scale of directed information flow. The raw-signal baseline model also benefited from pair-level aggregation (as can be seen in figure 5), achieving strong performance relative to lower aggregation scales. For Subject 5, the baseline achieved Pearson $r \approx 0.98$ and Spearman $\rho \approx 0.98$, with $R^2 \approx 0.95$. However, absolute error remained consistently higher compared to EPSTE (MAE ≈ 0.0034, RMSE ≈ 0.0042), indicating reduced precision in recovering STE magnitudes despite high correlation. This gap suggests that while aggregation alone recovers coarse directed structure, the polygonal symbolic representation enables more accurate encoding of causal strength.

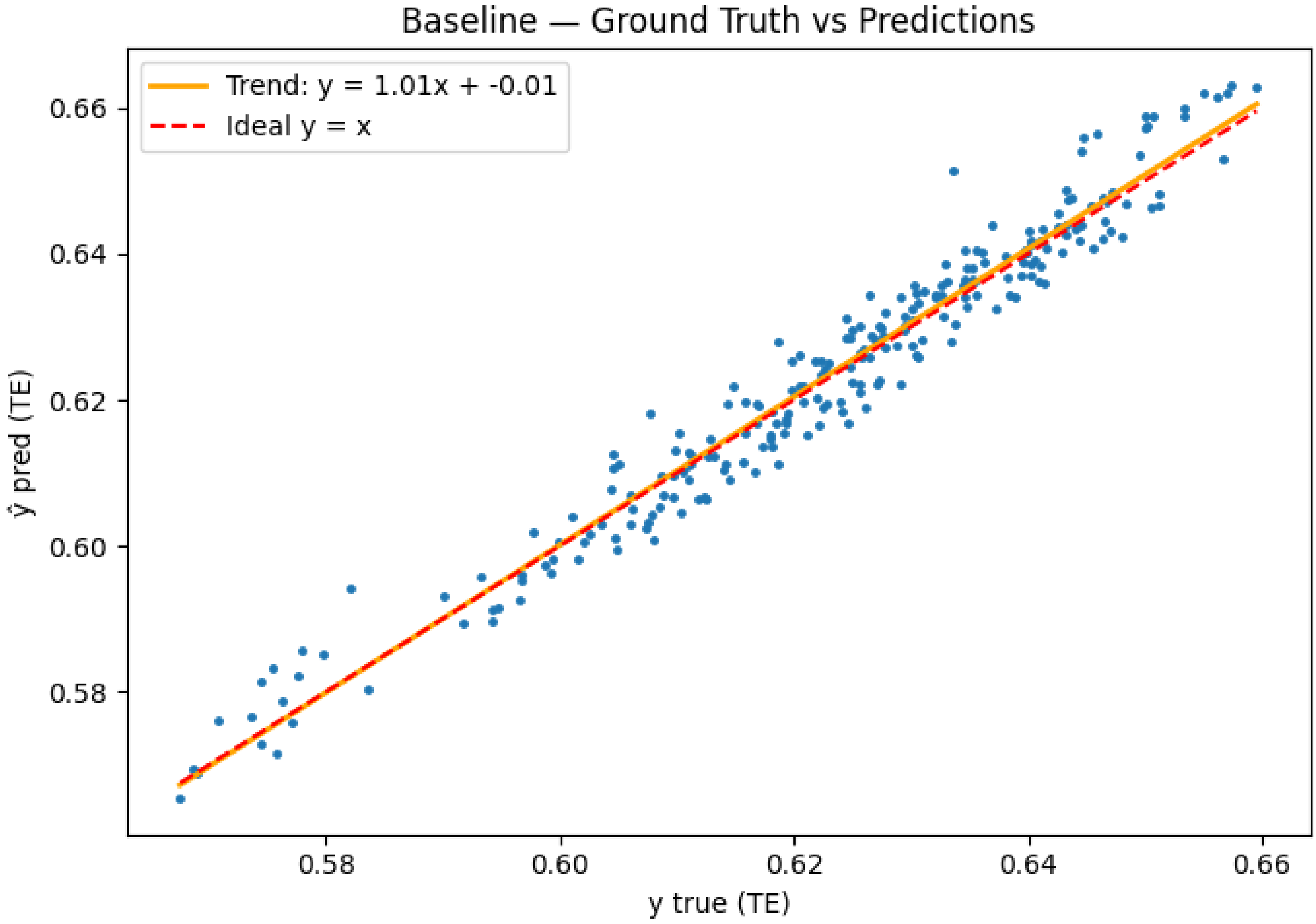

*Figure 5. Pair-level comparison between ground-truth transfer entropy values and predictions from the raw-signal baseline model for the held-out test subject. The dashed red line denotes the ideal identity relationship (y = x), while the fitted trend line indicates near-unity slope with slight residual scatter. Although aggregation yields strong linear correspondence, the broader dispersion relative to EPSTE reflects reduced precision in recovering fine-grained causal strength from raw symbolic amplitude representations.*

To assess whether the recovered pair-level predictions preserved interpretable directed structure beyond numerical accuracy, predicted Transfer Entropy values were visualised as directed connectivity heatmaps for the held-out test subject (Subject 5, LSD condition). Each heatmap represents the predicted directed influence from source channels (rows) to target channels (columns), allowing qualitative inspection of global structural organisation across the visual cortical network. The EPSTE-derived heatmap (figure 6) exhibited a coherent and varying structure, with consistent banding patterns and clear gradients of directed influence across channel pairs. Importantly, these patterns aligned closely with the structure implied by the ground-truth TE estimates, indicating that the triangle-based representation preserved not only pointwise accuracy but also the relational organisation of directed interactions across the network. The absence of spurious high-magnitude outliers and the continuity of values across neighbouring channel pairs further suggest that EPSTE captures stable, distributed causal structure rather than isolated pairwise artefacts.

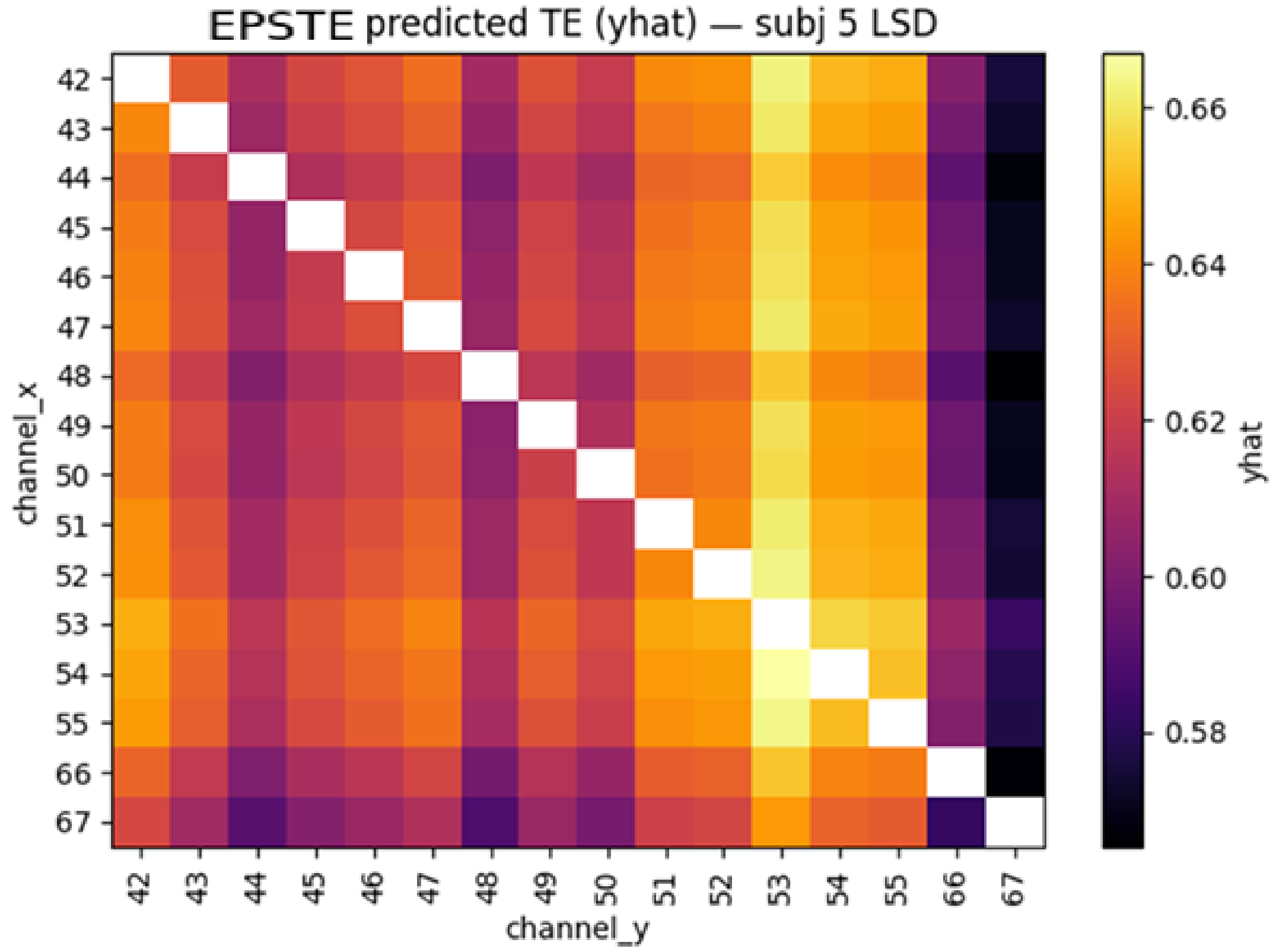

*Figure 6. Heatmap of pair-level transfer entropy predictions produced by the EPSTE model for the held-out test subject (Subject 5, LSD condition). Rows correspond to source channels (x) and columns to target channels (y); colour intensity indicates predicted directed information flow. The diagonal is masked (white) to exclude self-connections. The structured banding and graded variations across channel pairs indicate coherent, non-uniform directed connectivity, reflecting the EPSTE model's ability to preserve relational and directional structure beyond pointwise prediction accuracy.*

In contrast, while the baseline representation also produced a broadly interpretable heatmap (figure 7), its structure was comparatively compressed and less differentiated, with reduced contrast between dominant and weaker directed interactions. This visual attenuation is consistent with the lower pair-level accuracy metrics observed for the baseline model and suggests a diminished ability to resolve fine-grained directional structure despite using an identical neural architecture. Table 1 summarises the model performance metrics across window, trial, and pair aggregation levels for both the triangle-based and baseline representations. Notably, the triangle representation yields significantly lower error and higher correlation at the pair level, supporting its capacity to recover stable directed dependencies.

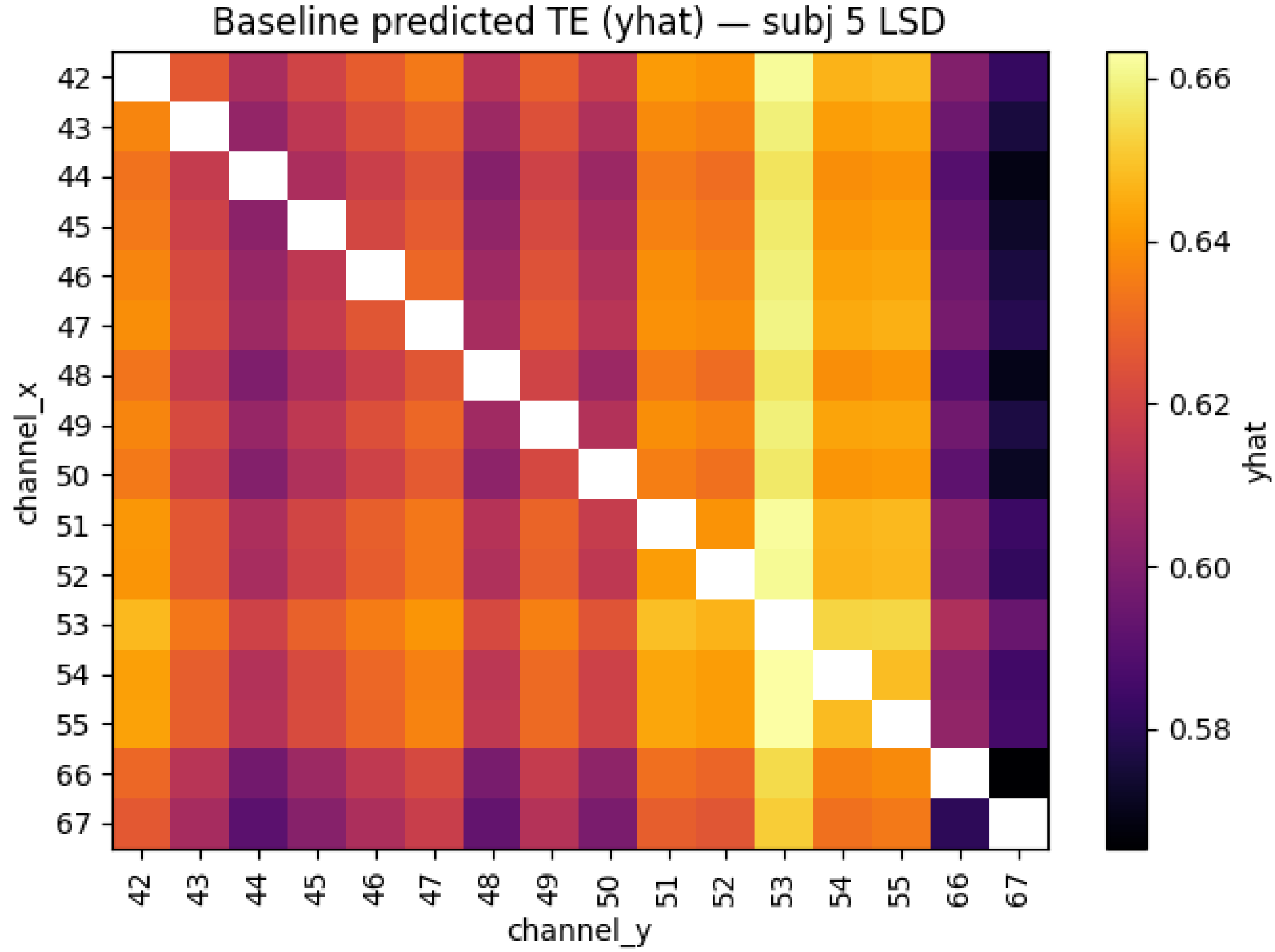

*Figure 7. Heatmap of pair-level transfer entropy predictions produced by the raw-signal baseline model for the held-out test subject (Subject 5, LSD condition). Rows denote source channels (x) and columns denote target channels (y), with colour intensity indicating predicted directed information flow; the diagonal is masked (white) to exclude self-connections. While the baseline recovers broad patterns of directed connectivity, the map appears more compressed with reduced contrast between stronger and weaker interactions compared to EPSTE, consistent with its higher absolute error and reduced ability to resolve fine-grained directional structure.*

**Table 1: Results Summary Table: Model Fit Metrics Across Aggregation Levels**

| *Aggregation Level* | *Representation* | *MAE* | *RMSE* | *Pearson r* | *Spearman ρ* | *$R^2$* |
|---|---|---|---|---|---|---|
| *Window-level* | Triangle | 1.0990 | 1.3624 | 0.4788 | 0.3996 | – 0.943 |
| | Amplitude Baseline | 0.0394 | 0.0505 | 0.5156 | 0.3961 | 0.097 |
| *Bag-level* | Triangle | 0.0214 | 0.0269 | 0.8622 | 0.8071 | 0.743 |
| | Amplitude Baseline | 0.0308 | 0.0390 | 0.6782 | 0.5495 | 0.460 |
| *Pair-level* | Triangle | 0.0021 | 0.0029 | 0.9933 | 0.9936 | 0.979 |
| | Amplitude Baseline | 0.0034 | 0.0042 | 0.9788 | 0.9792 | 0.954[1] |

*Table 1. Model performance metrics for triangle-based (EPSTE) and amplitude baseline representations across three aggregation levels: window, bag (trial-level), and pair (condition-level). Metrics include Mean Absolute Error (MAE), Root Mean Square Error (RMSE), Pearson correlation (r), Spearman rank correlation (ρ), and coefficient of determination ($R^2$). The triangle representation shows markedly improved pair-level fit, with higher correlation and reduced error, suggesting that geometric-symbolic representations better support the recovery of stable directed dependencies. A Wilcoxon signed-rank test confirmed that pair-level predictions from the triangle model significantly outperformed the baseline ($p = 2.9 \times 10^{-15}$).*

## 4.5 Surrogate Null Distribution and Significance Assessment

To assess whether observed Transfer Entropy (TE) estimates reflected genuine directed dependencies rather than chance temporal alignment, surrogate-based significance testing was performed at the pair level. For each directed channel pair in the held-out test subject (Subject 5, LSD condition), a null distribution was generated using phase-randomised surrogates that preserved the marginal power spectrum and autocorrelation structure while destroying directed temporal dependence. Observed TE values were then normalised relative to their surrogate distributions, yielding z-scores and empirical p-values. Figure 8 shows the distribution of surrogate-normalised TE z-scores across 200 randomly sampled channel pairs. The distribution is centred near zero, as expected under the null hypothesis of no directed interaction, with variance determined by the finite surrogate sample size ($N = 25$). Most channel pairs fall within ±2 standard deviations of the null mean, indicating no statistically significant directed influence at the chosen threshold ($p < 0.05$). A small number of edges exhibit larger positive or negative deviations, consistent with weak but detectable departures from the null.

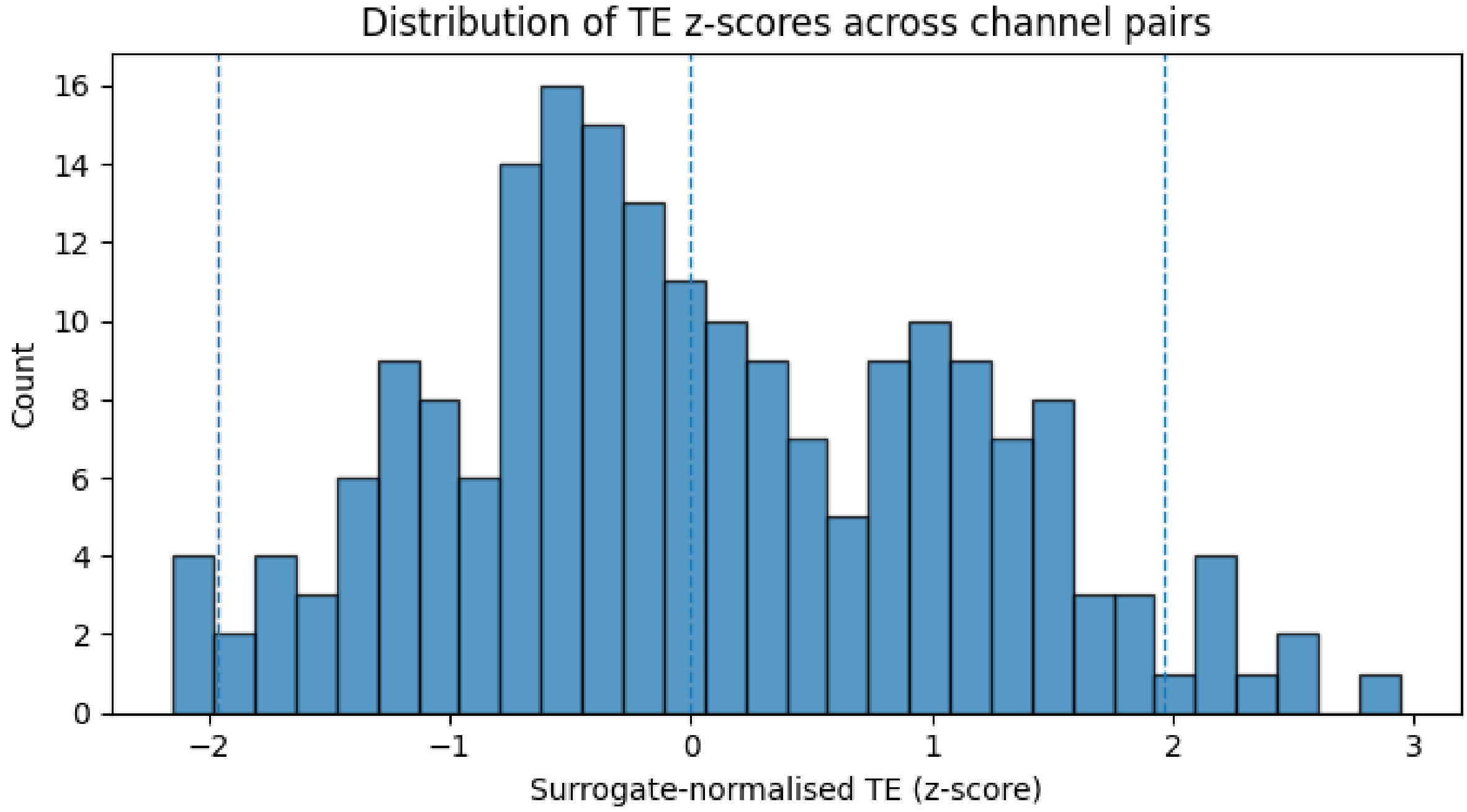

*Figure 8. Histogram of surrogate-normalised TE z-scores for 200 channel pairs from the held-out test subject (Subject 5). Vertical dashed lines indicate the null mean (z = 0) and ±2 standard deviations. The distribution is centred near zero, as expected under the null hypothesis of no directed interaction, with most edges falling within the null bounds and a small subset exhibiting deviations consistent with statistically detectable directed influence beyond chance.*

Importantly, this surrogate analysis serves a dual role. First, it validates that raw TE estimates are not trivially inflated by autocorrelation or shared spectral structure. Second, it provides statistically grounded target values for training and evaluating EPSTE, ensuring that learned mappings reflect directed information rather than spurious correlations. The surrogate-normalised TE framework therefore underpins both the statistical validity and the supervision strategy employed in this study. As can be observed from the histogram, supervision is weak and constitutes a fundamental limitation to this study.

## 4.6 Statistical Comparison and Error Structure

To formally assess whether the observed performance differences between representations reflect robust improvements rather than chance variation, pair-level prediction errors were compared directly between the EPSTE and raw-signal baseline models for the held-out test subject (Subject 5). All comparisons were performed on a per-edge basis, preserving paired correspondence between methods and ensuring statistical validity.

At the pair level, EPSTE consistently yielded lower absolute error than the baseline across directed edges. The mean absolute error for the baseline model was MAE ≈ 0.00340, compared to MAE ≈ 0.00209 for EPSTE, corresponding to a mean absolute improvement (baseline − EPSTE) of approximately $1.31 \times 10^{-3}$. This reduction is substantial relative to the overall scale of the TE values being estimated. Crucially, the improvement was not driven by a small number of extreme edges. The median absolute error improvement was $\approx 1.16 \times 10^{-3}$, with an interquartile range of approximately ($-3.1 \times 10^{-4}$, $2.66 \times 10^{-3}$). While a minority of edges showed comparable or slightly worse performance under EPSTE, the predominance of positive improvements indicates that EPSTE outperformed the baseline across the majority of directed pairs. This distributional structure is reflected in the improvement histogram, where the mass of the distribution lies clearly above zero. A paired Wilcoxon signed-rank test was used to assess statistical significance, as error distributions were non-Gaussian and paired by construction. The test revealed an extremely significant difference between methods ($p \approx 2.9 \times 10^{-15}$), providing strong evidence that the observed reduction in absolute error achieved by EPSTE is not attributable to random variation.

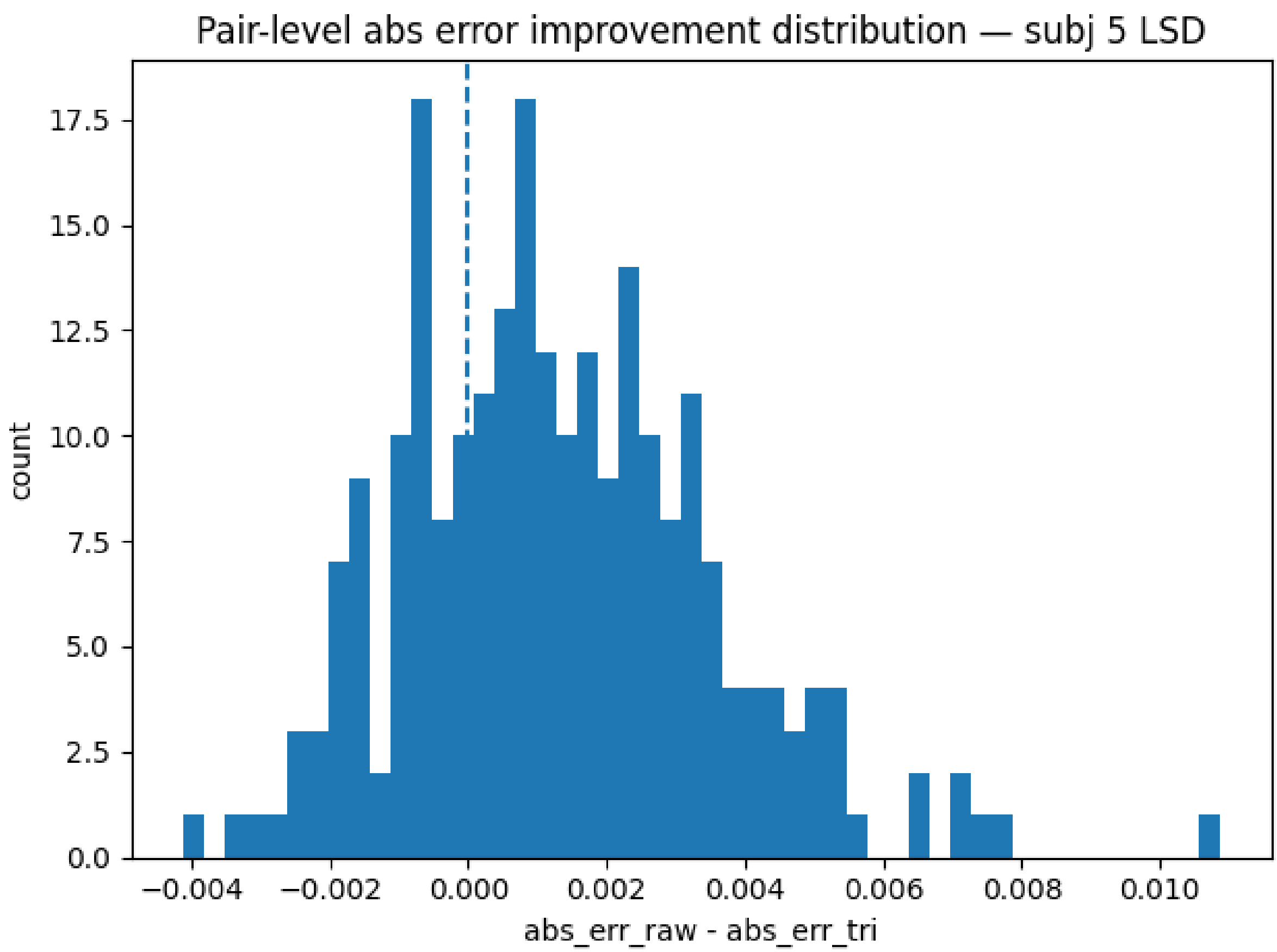

*Figure 9. Histogram of pair-level absolute error differences for the held-out test subject (Subject 5, LSD condition), computed as* $|error_{baseline}|$ - $|error_{EPSTE}|$ *for each directed channel pair. Positive values indicate lower error under the EPSTE representation. The dashed vertical line denotes zero improvement and distribution is predominantly shifted toward positive values, indicating that EPSTE achieves lower absolute error than the baseline for many directed pairs, with improvements not driven by a small number of extreme edges but distributed broadly across the network.*

Beyond central tendency, error dispersion across trials was examined to assess the stability of inferred causal estimates. EPSTE exhibited a higher median edgewise standard deviation across trials (≈ 0.040) compared to the baseline (≈ 0.028). When normalised by mean edge strength using the coefficient of variation (CV), EPSTE again showed higher median values (≈ 0.064 vs ≈ 0.045).

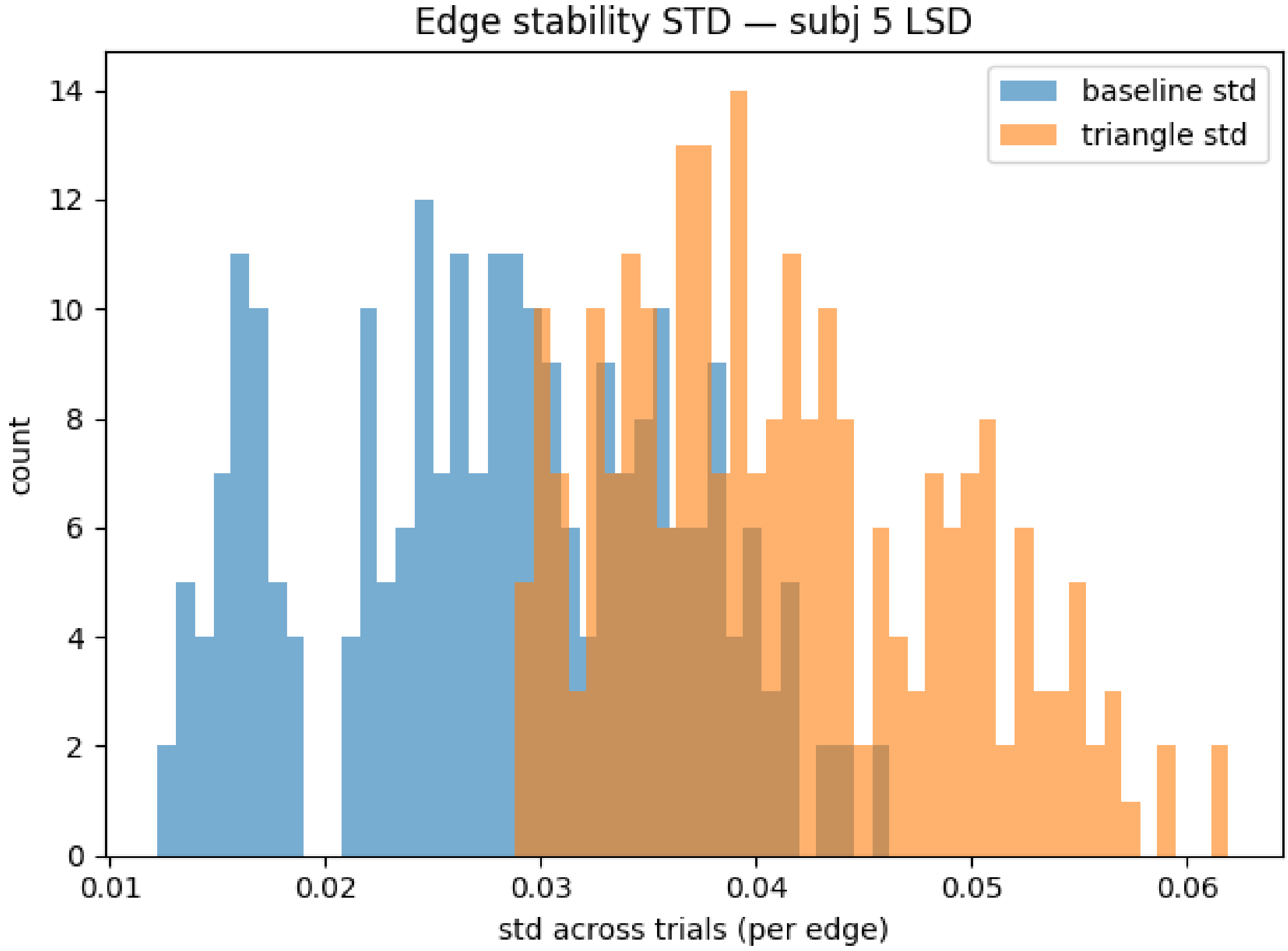

*Figure 10. Histogram of per-edge standard deviation of predicted Transfer Entropy across trials for the held-out test subject (Subject 5, LSD condition). Blue bars show the raw-signal baseline, while orange bars show the triangle-based EPSTE representation. EPSTE exhibits higher edgewise variability, reflecting a broader dynamic range and greater sensitivity to trial-level fluctuations, whereas the baseline distribution is more compressed. In the context of substantially lower absolute error and higher variance explained for EPSTE, this increased dispersion is interpreted as responsiveness to genuine trial-to-trial variation rather than reduced reliability.*

However, when normalised ( as can be observed in figure 11) by the mean edge strength using the coefficient of variation (CV), EPSTE again showed higher median values (≈ 0.063 vs ≈ 0.046).

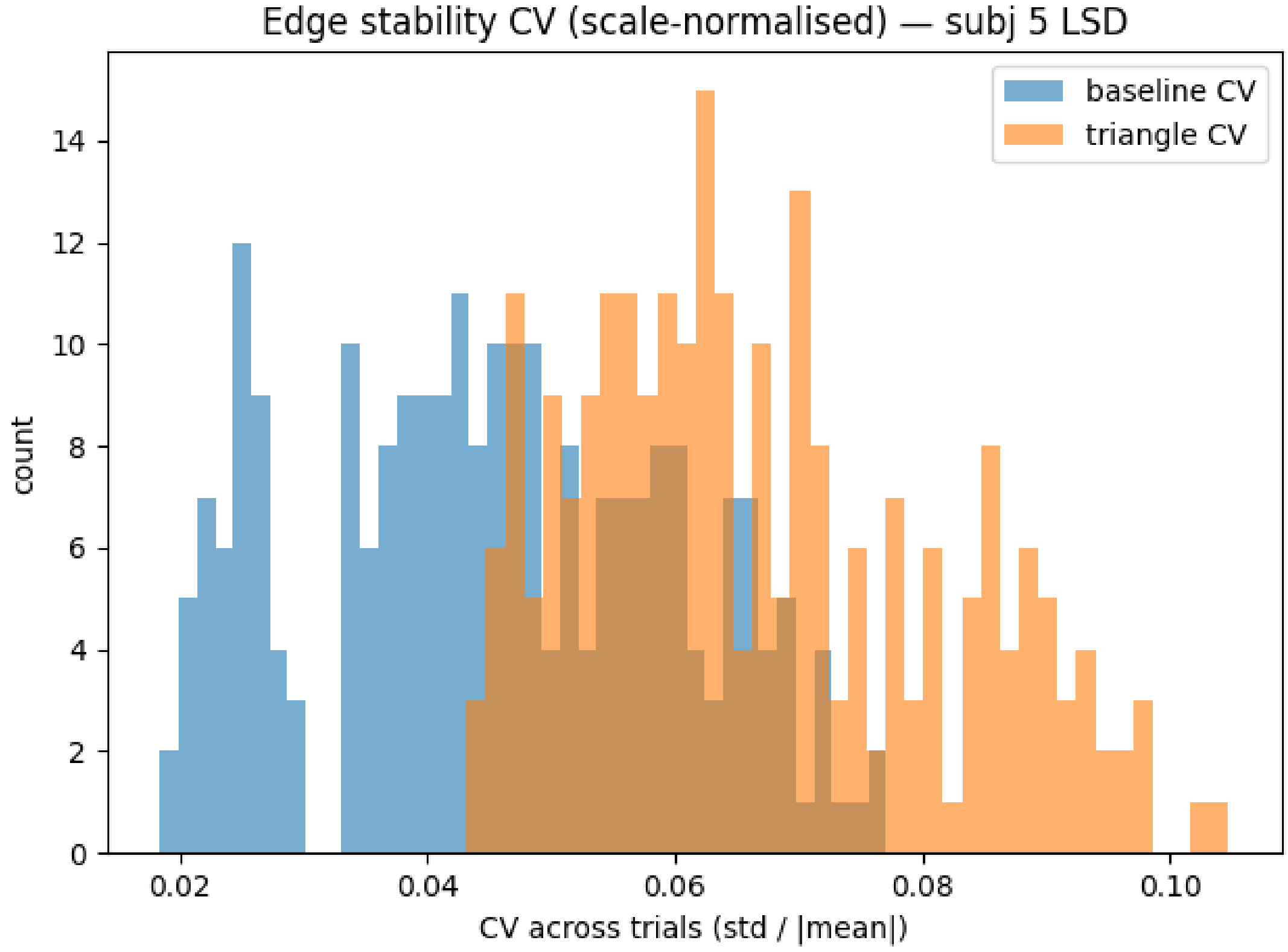

*Figure 11. Histogram of the coefficient of variation (CV = std / |mean|) of predicted Transfer Entropy across trials for each directed edge in the held-out test subject (Subject 5, LSD condition). Blue bars show the raw-signal baseline and orange bars the triangle-based EPSTE representation. EPSTE exhibits higher CV values, indicating greater relative variability after normalisation by edge strength. In conjunction with EPSTE's lower absolute error and higher variance explained, this pattern suggests increased sensitivity to meaningful trial-level fluctuations rather than instability, whereas the baseline shows compressed variability consistent with attenuation of dynamic structure.*

Importantly, this increased dispersion should not be interpreted as reduced reliability. Rather, it reflects the greater dynamic range and sensitivity of the EPSTE representation. Because EPSTE recovers TE values with substantially lower absolute error and higher variance explained ($R^2 \approx 0.98$) than the baseline ($R^2 \approx 0.95$), increased variability across trials is consistent with responsiveness to genuine trial-level fluctuations rather than noise amplification. In contrast, the lower dispersion observed in the baseline model is consistent with systematic compression toward the mean, a pattern also reflected in its higher absolute error and reduced variance explained. Table 2 summarises key error and variability metrics at the pair level, highlighting that the triangle-based representation significantly outperforms the baseline in absolute prediction error, despite slightly increased variability across samples. Taken together, the convergence of multiple statistical indicators, lower mean and median absolute error, predominantly positive improvement distribution, an extremely significant paired non-parametric test, and superior variance explanation, provides a strong and defensible basis for concluding that EPSTE yields more accurate and informative pair-level causal estimates than the raw-signal baseline.

**Summary of Pair-Level Prediction Error and Variability Metrics**

| Metric | Baseline | Triangle | Difference (Baseline / Triangle) |
|---|---|---|---|
| Median Std. Dev. of Predictions | 0.0283 | 0.0400 | – |
| Median Coefficient of Variation (CV) | 0.0453 | 0.0636 | – |
| Mean Absolute Error (MAE) | 0.00340 | 0.00209 | 0.00131 |
| Median Absolute Error | – | – | 0.00116 |
| Interquartile Range (IQR) of Improvement | – | – | (–0.00031, 0.00266) |
| Wilcoxon Signed-Rank Test (p-value) | – | – | $2.91 \times 10^{-15}$ |

*Table 2. Pair-level summary statistics comparing prediction variability and absolute error between the baseline (raw amplitude) and triangle-based symbolic representations. The triangle representation yielded lower mean absolute error, with a median improvement of ~0.00116 and a highly significant advantage confirmed by a Wilcoxon signed-rank test ($p < 10^{-14}$). Variability (CV, std) is slightly higher in triangle representations due to preserved geometric variation.*

## 4.7 Model Training and Performance

The raw-signal baseline exhibited steady but shallow convergence. Training loss decreased from approximately 0.29 at epoch 1 to ≈ 0.20 by epoch 30–35, after which improvements plateaued. Validation loss closely tracked training loss throughout optimisation, stabilising around ≈ 0.24. The close alignment between training and validation curves indicates that the baseline model did not substantially overfit; however, the relatively high loss floor suggests limited representational capacity to capture the relevant causal structure from raw amplitude features alone. On the held-out test subject, baseline performance improved with aggregation but remained comparatively constrained. At the window level, prediction error was high (MSE ≈ 0.86) despite a moderate Pearson correlation (r ≈ 0.52), indicating noisy local predictions with limited explanatory power. Aggregation at the bag level reduced error (MSE ≈ 0.52) and increased correlation (r ≈ 0.68), but substantial variance remained unexplained. At the pair level, performance improved markedly (MSE ≈ 0.0061, r ≈ 0.98), demonstrating that aggregation alone enables partial recovery of directed dependencies, albeit with persistent systematic error.

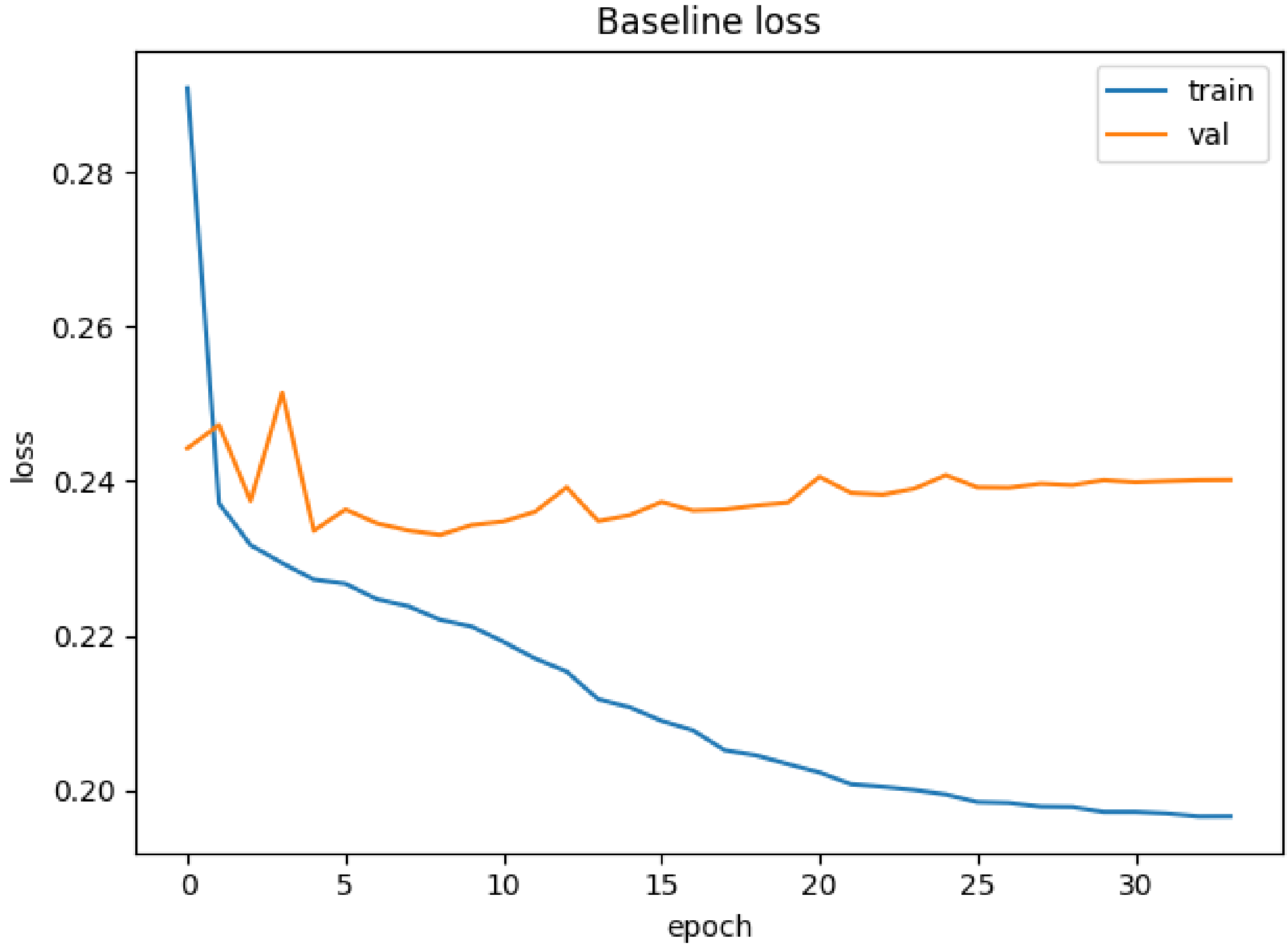

*Figure 12. Training and validation loss curves for the raw-signal baseline model. Loss is shown as mean squared error (MSE) over training epochs. Training loss decreases steadily, while validation loss plateaus early and remains stable, indicating convergence without strong overfitting.*

In contrast, the EPSTE model converged more rapidly and to a substantially lower loss regime (as can be observed in figure 13). Training loss decreased from ≈ 0.20 at epoch 1 to ≈ 0.10 by epoch 50, while validation loss stabilised around ≈ 0.12. Although validation loss exhibited mild oscillations during early training, it remained consistently well below that of the baseline model, indicating improved generalisation rather than overfitting. The sustained separation between baseline and EPSTE loss curves reflects the greater informativeness of the symbolic geometric representation, which enables the network to extract structured dependencies rather than fitting noise. Test-set performance further highlights this distinction. As expected, window-level EPSTE predictions remained noisy (MSE ≈ 1.86, r ≈ 0.48), reinforcing the interpretation that local windows do not encode stable causal structure in isolation. Again, aggregation produced substantial gains. At the bag level, EPSTE achieved markedly lower error (MSE ≈ 0.25) and stronger correlation with ground-truth TE (r ≈ 0.86) than the baseline. At the pair level, EPSTE achieved very low error (MSE ≈ 0.0028) and near-perfect correspondence with ground truth (r ≈ 0.993).

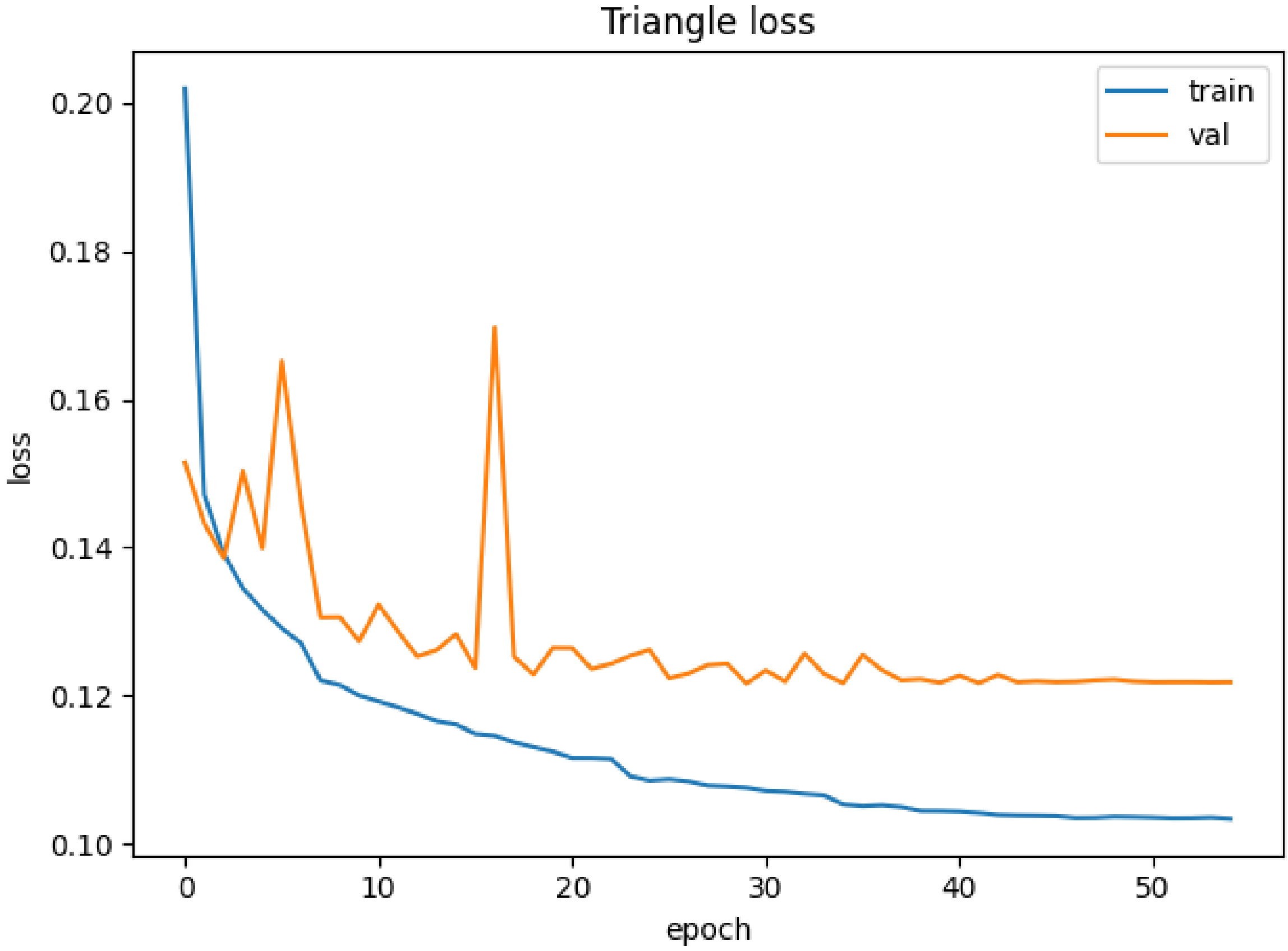

*Figure 13. Learning dynamics of the EPSTE model across training epochs, showing training loss (blue) and validation loss (orange). The model converges rapidly from a higher initial loss to a substantially lower loss regime than the baseline, with validation loss stabilising around ≈ 0.12 after early transient oscillations. The sustained separation between training and validation curves indicates effective generalisation without pronounced overfitting, while the lower loss floor reflects the increased informativeness of the geometric–symbolic representation in supporting stable learning of directed dependencies. The transient spike in validation loss around epoch 16 likely reflects a single validation batch that was particularly difficult for the model to reconcile, rather than a systematic instability.*

# 5. Discussion

## 5.1 Summary of Key Findings

The above results show key findings; most importantly a Wilcoxon signed-rank test was used to assess whether the triangle-based representation (EPSTE) yielded significantly lower pairwise prediction errors than the baseline. The resulting p-value ($p < 10^{-14}$) allowed rejection of the null hypothesis: *"Polygon-based symbolic encoding does not increase learnability more effectively than classical amplitude-based time series representations"* and support the claim that *"Polygon-based symbolic encoding increases learnability more effectively than classical amplitude-based time series representations".*

Importantly this study makes three core contributions: (1) it introduces a symbolic triangle-based representation (EPSTE) that enhances symbolic transfer entropy estimation from neural time series, (2) it demonstrates that representational structure can outperform raw signal amplitude in learning directed dependencies, and (3) it proposes a generalisable framework for symbolic geometry in causal inference, with potential relevance for neural decoding tasks. The results demonstrate that, fundamentally, aggregation plays a functional and necessary role in enabling representation learning within this framework. Local window-level predictions are noisy and weakly informative in isolation. However, their integration at the trial (bag) level allows the model to begin recovering stable directed dependencies, with further aggregation to the pair (edge) level yielding the strongest and most interpretable causal estimates. The behaviour is expected under the experimental framing adopted in this work. Individual windows capture only short, local segments of neural activity and are not assumed to encode stable directed dependence in isolation. This behaviour aligns with the Law of Large Numbers and supports the interpretation that causal structure in neural systems is distributed and might only become reliably observable at appropriate scales of aggregation. Transfer Entropy reflects statistical dependencies accumulated across time, and its estimation from local observations is inherently noisy. Consequently, window-level outputs are not interpreted as direct causal estimates.

Instead, window-level predictions are treated as a diagnostic representation of model behaviour, providing local contributions that support higher-order aggregation. Under this interpretation, weak or unstable window-level performance is a natural consequence of attempting to infer distributed causal structure from noisy, nonlinear neural interactions, rather than evidence of modelling failure. The critical question is therefore not whether causal structure is recoverable at the window level, but whether aggregation enables the emergence of real and trustworthy causality at higher scales vs a statistical artifact of the aggregation method.

Crucially, the EPSTE-based representation yields lower absolute error, statistically significant improvements, and robust gains across median and interquartile ranges, rather than effects driven solely by mean performance. Because both representations were evaluated using identical architectures, optimisation procedures, and supervision, these improvements are attributable to representational structure rather than model capacity or overfitting. The convergence of multiple evaluation criteria, including absolute error, dispersion measures, rank-based correlations, and non-parametric significance testing, suggesting a reasonably defensible basis for concluding that increased representational geometry, combined with appropriate aggregation, more effectively supports the recovery of stable directed causal structure at the pair level. Beyond aggregate performance metrics, multiple diagnostic analyses provide convergent evidence that the observed improvements arise from meaningful representation learning rather than optimisation artefacts. Training and validation loss curves reveal clear differences in learning dynamics between representations. The raw-signal baseline exhibited shallow convergence, with training and validation loss stabilising early at a relatively high loss floor. Although the close alignment between training and validation curves indicates limited overfitting, the early plateau suggests restricted representational capacity to encode directed dependencies from raw amplitudes alone. The EPSTE model converged more rapidly and to a substantially lower loss regime, with validation loss consistently remaining well below that of the baseline. This sustained separation indicates improved generalisation and continued extraction of informative structure rather than memorisation of noise. Second, structural correspondence between predicted and ground-truth transfer entropy was examined using pair-level heatmaps. EPSTE heatmaps exhibited sharper contrast, clearer edge differentiation, and stronger alignment with the ground-truth connectivity structure, whereas baseline heatmaps showed compression toward the mean and reduced dynamic range. These qualitative differences are consistent with the quantitative error and variance-explained results and suggest that geometric symbolic encoding preserves relational structure that is obscured in raw-signal representations.

Together, learning curves and empirically derived visualisation, provide complementary evidence that the performance gains observed for EPSTE reflect genuine representational advantages rather than optimisation artefacts or statistical bias. Under identical neural architectures, optimisation schedules, and supervision, the EPSTE representation yielded significantly lower pair-level prediction error than the standard symbolic baseline with improvements observed across mean, median, and interquartile ranges. A paired non-parametric Wilcoxon signed-rank test revealed an extremely significant difference in absolute error distributions ($p \ll 0.001$) indicating that the observed performance gains are highly unlikely to arise from random variation alone. However, this rejection applies specifically to the effect of representational choice as implemented in this study and does not isolate geometric primitives independently of binning discretisation. Accordingly, the results support the conclusion that the EPSTE representation when taken as a complete symbolisation strategy enables more accurate recovery of directed dependencies, while leaving open the question of the degree of contribution of geometry versus bin-induced redistribution of information.

## 5.2 General Interpretation and Impact

A central challenge in applying Transfer Entropy (TE) to empirical neural data lies in the curse of dimensionality inherent in the conditional mutual information estimation. Classical TE estimators require reliable estimation of joint probability distributions over past and future states of interacting processes. As temporal embedding depth or conditioning order increases, the effective state space grows exponentially and rapidly inflates data requirements and estimator variance. These limitations are well documented in the literature, particularly for finite-sample neural data where trial counts and recording durations are constrained (Paninski, 2003).

The results of the present study align closely with these theoretical concerns. At the window level, both TE supervision and model predictions exhibited high variance and poor explanatory power which reflects the instability of local estimates derived from sparse sampling of high-dimensional state spaces. Importantly, this instability was not resolved by the learning architecture alone, but was progressively reduced through aggregation across time and trials. This behaviour is consistent with the view that reliable directional structure in neural systems emerges only after sufficient accumulation of evidence rather than being recoverable from short local segments. Symbolic formulations of TE were introduced specifically to mitigate these estimation challenges, by discretising continuous signals into symbolic sequences, methods such as Symbolic Transfer Entropy (STE) and ordinal or phase-based variants reduce estimator variance and improve robustness to noise. However, this robustness is typically achieved through aggressive compression, collapsing amplitude, curvature, and waveform morphology into low-cardinality alphabets. Ordinal-only encodings, for example, preserve relative ordering while discarding absolute magnitude and local shape information (Dimitriadis et al., 2016).

The present findings empirically support this trade-off. While symbolic abstraction is benefited from tractable TE estimation, overly coarse symbolisations risk representational impoverishment. The improved performance observed for EPSTE suggests that symbolic representations which preserve richer local geometric structure can better balance variance reduction with informational expressiveness, without increasing temporal conditioning order or incurring the exponential growth in estimator dimensionality associated with deeper embeddings.

## 5.3 Neural Approximations of Transfer Entropy

Most neural TE estimators operate directly on raw continuous signals, implicitly assuming that representation choice is secondary to model capacity (Shorten et al., 2020). The present results challenge the assumptions and magnitude of usefulness of raw continuous signals. Despite identical architectures, optimisation procedures, and supervision, substantial performance differences emerged solely due to representational choice. This indicates that neural models do not eliminate the need for appropriate representations; rather, they shift the burden of estimation onto the structure of the input space. Taken together, these findings support the view that representational structure plays a critical role in the practical learnability of information-theoretic dependencies. In this sense, representation precedes learning: the capacity of neural estimators to recover directed information flow is fundamentally constrained by how temporal structure is encoded before learning begins.

## 5.4 Representational Geometry as an Enabling Factor

The central contribution of this work lies in demonstrating that representational geometry can function as an enabling factor for learning information-theoretic dependencies, rather than merely as a preprocessing convenience. The EPSTE framework does not increase temporal conditioning order or embedding depth in the conventional sense. Instead, it redistributes information contained in local temporal structure across multiple interpretable geometric dimensions, thereby improving the efficiency with which causal dependencies can be learned.

In the proposed representation, local triplets of samples capture complementary aspects of local signal morphology that are collapsed in raw amplitude representations and largely discarded in ordinal-only symbolic encodings. By distributing information across multiple geometric attributes, the representation increases separability between local states without requiring higher-dimensional temporal embeddings, helping to mitigate the curse of dimensionality that constrains classical TE estimation. By embedding richer relational information into each symbolic unit, the representation supports greater variability and niche formation within the state space. Small differences in waveform shape, such as changes in curvature or directional trend that would otherwise be indistinguishable under amplitude- or order-based encodings map to distinct symbolic configurations. This expanded internal structure allows the learning system to resolve subtle but causally relevant distinctions, supporting more stable and discriminative learning. In this sense, representational geometry does not merely reduce variance; it enriches the descriptive capacity of the symbolic space, allowing variation to be expressed, preserved, and exploited rather than averaged away (Liang et al., 2025). As in linguistic or symbolic reasoning systems, increased internal structure enables more expressive modelling while maintaining tractability, providing a scaffold on which higher-order dependencies can be learned efficiently (Besold et al., 2021).

An instructive analogy can be drawn from speech and language processing. Human listeners do not infer meaning or structure in language directly from the raw amplitudes of acoustic waveforms. Instead, auditory signals are transformed into structured symbolic units, such as phonemes and syllables, that encode relational patterns in the time and frequency domain (Casserly & Pisoni, 2010; Mai et al., 2024). These symbolic units provide a syntactic scaffold that allows higher-level dependencies, such as grammatical structure and semanticity, to be inferred efficiently and robustly, even in the presence of noise and variability (Bohn et al., 2020). From an information-theoretic perspective, this geometric encoding concentrates predictive information into a compact but expressive state space. This reduces estimator burden, stabilises learning dynamics, and lowers loss floors during training, as observed empirically in this work. Importantly, this work does not claim that triangular primitives are optimal or universal. Rather, the results support the broader principle that appropriately structured representations can substantially enhance the practical learnability of causal dependencies by aligning the representation with the statistical and morphological structure of the underlying signal.

## 5.5 Broader Implications for Causal Inference in Neuroimaging

The findings of this study suggest that causal inference in neuroimaging should not be treated as an estimator-only problem. While considerable effort has focused on improving statistical estimators of directed dependence, the present results indicate that representational choice is a first-order methodological decision that strongly conditions what causal structure can be learned in practice. Even when identical learning architectures and supervision are used, differences in representation produce substantial differences in stability, accuracy, and interpretability.

More broadly, the use of symbolic–geometric representations highlights a general strategy for managing complexity in high-dimensional, noisy dynamical systems. By embedding structured local information into compact symbolic units, such approaches may help stabilise causal inference across modalities where data are limited or nonstationary, including fMRI, behavioural time series, and other complex economic, biological or physical systems. These implications are methodological rather than domain-specific, emphasising representation as a critical component of practical causal analysis across domains.

## 5.6 Limitations

Several methodological limitations should be acknowledged. First, EPSTE depends on symbolic binning choices, including the number of bins and the quantile thresholds used to discretise geometric features. Although quantile binning improves robustness to scale changes and outliers, alternative discretisation strategies (e.g., adaptive binning or learned vector quantisation) could alter the symbolic dynamics and therefore influence both TE targets and learned mappings (Barà et al., 2023). Because EPSTE and the raw-signal baseline operate on different feature spaces, their empirical symbol distributions differ by construction. As a result, it cannot be fully disentangled whether performance differences arise from the geometric triangular representation itself or from how information is redistributed across symbolic bins.

Second, the symbolic abstraction necessarily discards some fine-grained amplitude information. While the triangle-based representation preserves local morphology via complementary geometric features (e.g., magnitude and curvature proxies), it does not retain the full continuous waveform. Consequently, effects that are strongly amplitude- or frequency-specific may be partially attenuated, particularly when informative structure is not well captured by the chosen primitive set.

Third, the framework relies on surrogate-based supervision to provide grounded target values for learning. Phase-randomised surrogates' control for autocorrelation and shared spectral structure, but they do not constitute a mechanistic "ground truth" causal model. Unfortunately, most of the TE values used to compute the ground truth for the network did not reach significant difference from the null distribution, rendering the ground truth and supervision relatively weak targets. Accordingly, a learned predictor should be interpreted as approximating *a specific operational definition* of TE under this surrogate-controlled estimation.

Fourth, and perhaps most importantly, this study concerns the selection of temporal lags used for Transfer Entropy estimation. Lag values were constrained to a physiologically plausible range reflecting known neural propagation delays. While this grounds the analysis in biological realism, it restricts the temporal scope of detectable interactions and may underrepresent slower or heterogeneous dependencies such as modulatory or cross-frequency effects. Rather than being inferred from the data, lag selection was treated as a fixed hyperparameter. Although a maximum-over-lags strategy reduces sensitivity to individual lag choice, it assumes that relevant causal interactions fall within the predefined range. This limitation reflects a broader challenge in practical TE estimation, where expanding lag ranges rapidly increases estimator variance and computational cost (Vicente & Wibral, 2014). Future work could address this by learning lag distributions directly or jointly optimising lag selection alongside symbolic representation and model parameters.

Finally, although this study operates in source space using AAL90-parcellated time series, limitations remain. Source reconstruction reduces sensor-level mixing, but it does not eliminate leakage or shared source components entirely, and residual spatial cross-talk may still affect directionality estimates. Thus, inferred edges should be interpreted as directed *functional* influences between AAL regions under the chosen preprocessing and reconstruction pipeline, rather than definitive anatomical pathways.

## 5.7 Nature of the “Causality” Inferred

It is important to clarify the nature of the “causality” inferred in this work. Transfer Entropy, and by extension EPSTE, quantifies directed statistical dependence and predictive information flow rather than mechanistic or interventional causation. Directionality reflects whether past states of one process reduce uncertainty about another’s future beyond the target’s own history, not whether one region exerts a direct physical force on another and accordingly, the causal relationships identified should be interpreted as informational or functional influences within a complex dynamical highly interactive system. This framing avoids overinterpretation while remaining consistent with established causal-inference practice in non-invasive neuroimaging, where experimental interventions are typically absent and causal claims must remain explicitly conditional on modelling and estimation choices (Chiarion et al., 2023).

## 5.8 Future Directions

The EPSTE framework motivates several natural extensions. A first direction is systematic optimisation of design choices such as lag sets, bin counts, and network hyperparameters. Evolutionary optimisation (e.g., genetic algorithms or NEAT-style search) could be used to explore this high-dimensional configuration space and test whether performance gains persist across broader physiological lag ranges and alternative symbolic vocabularies.

A second direction concerns representational generalisation. While triangles provide a compact and interpretable primitive, richer families of geometric descriptors (e.g., multiscale curvature, piecewise-polynomial segments, or mixed primitive sets) could capture complementary temporal motifs. Likewise, learning the symbolisation stage rather than fixing bin edges a priori could allow the model to discover a task-adaptive symbolic code that maximises predictability and separability.

A third extension is to multivariate or conditional TE, where causal influence is assessed while conditioning on additional processes. This would enable stronger controls for indirect pathways and shared drivers, and would align more closely with network-level causal interpretation, though it increases statistical and computational demands.

Finally, the broader idea suggested by EPSTE is that time series may admit a learnable "syntax": structured motifs function as discrete tokens whose ordering encodes higher-level dependencies. This invites a speculative but testable direction: training generative sequence models on symbolic motif streams (e.g., triangle tokens) in a self-supervised manner, analogous to large language modelling in natural language processing. Instead of predicting words, the model would predict future motif tokens, learning a latent "neural grammar" of temporal organisation. Such a model could support generative simulation, anomaly detection, or representation learning for downstream causal inference—provided it is framed as modelling statistical structure in symbolic dynamics rather than "decoding thoughts" or claiming mechanistic neural code. More broadly, learning symbolic temporal grammars that encode directed dependencies may offer a route toward generative models that respect physical, causal and dynamical constraints, addressing a recognised limitation of current video and simulation-based generative systems, which often produce physically implausible events due to the absence of explicit representations of causal structure and temporal dependency (Gupta et al., 2024).

# 6. Conclusion

This work set out to examine whether the practical estimation of Transfer Entropy from neural time series can be improved by altering how temporal structure is represented prior to learning. The results demonstrate that aggregation is essential: directed causal structure is not reliably observable at the level of short local windows but emerges through integration across trials and channel pairs. Within this multiscale framework, representational choice proves decisive. By encoding local temporal morphology into structured geometric symbols, EPSTE redistributes predictive information across interpretable dimensions without increasing temporal conditioning order. This enables more stable learning, lower estimation error, and improved recovery of directed dependencies compared to a raw symbolic baseline trained under identical conditions. These gains are therefore attributable to representational structure rather than model capacity or optimisation.

Importantly, this study does not claim to recover mechanistic causality. Instead, it demonstrates that informational or predictive causal structure becomes more learnable when the representation aligns with the statistical and morphological structure of the underlying signal. Taken together, the findings support a broader methodological conclusion: in data-limited, noisy settings such as neuroimaging, causal inference is fundamentally constrained not only by estimators, but by the language in which temporal dynamics are described. Representation, learning, and causality are inseparable components of effective inference.